\documentclass{aa}
\usepackage{graphicx}
\usepackage{txfonts}
\usepackage{xcolor}
\usepackage{soul}
\graphicspath{ {figures/} }

\def\Gaia{\textit{Gaia}\xspace}

\begin{document} 

\title{Infrared period--luminosity relations of Galactic Miras based on multi-epoch photometry and the \Gaia parallax uncertainty}

\author{S.~Uttenthaler\inst{\ref{inst_iap},\ref{UoV}}\fnmsep\thanks{Corresponding author: S. Uttenthaler,\\ {\tt stefan.uttenthaler@gmail.com}}
\and
T.~Lebzelter\inst{\ref{UoV}}
\and
S.~Meingast\inst{\ref{UoV}}
}

\institute{
Institute of Applied Physics, TU Wien, Wiedner Hauptstra\ss e 8-10, 1040 Vienna, Austria\label{inst_iap}
\and
Department of Astrophysics, University of Vienna, T\"urkenschanzstra\ss e 17, 1180 Vienna, Austria\label{UoV}
}

\date{Received February 16, 2026; accepted July 25, 2026}

\abstract
{Miras and other long-period variable (LPV) stars on the asymptotic giant branch (AGB) follow period--luminosity (PL) relations. These relations have been difficult to study for Galactic LPVs because their distances were poorly known in the past.}
{We aimed to establish the PL relations of solar neighbourhood Miras for several near-IR photometric bands.}
{To this end, we used multi-epoch photometry from the DIRBE and unTimely/WISE catalogues, \Gaia parallax distances, and contemporary pulsation periods obtained from optical observations of a well-selected sample of solar neighbourhood Miras.}
{We show that clearly defined PL relations in the nine investigated near-IR bands emerge from our data, for which we report slopes and zero-point magnitudes. We find that the Galactic Miras are fainter in the $K$ band than their siblings in the Large Magellanic Cloud. We derived the average period--temperature, period--bolometric-luminosity, and period--radius relations from fits to synthetic spectral energy distributions constructed from the PL relations. By applying AGB evolutionary models, the scatter of stars around the PL sequences can also be used to test whether the parallax uncertainties quoted in the \Gaia catalogue are realistic or underestimated. Furthermore, we performed such tests based on a comparison with VLBI parallaxes and with a sample of LPVs in the globular cluster 47~Tuc.}
{We conclude that for Galactic Miras with a fractional parallax uncertainty of $\le0.1$ in the \Gaia catalogue, the parallax uncertainty is underestimated by factors between 1.0 and 1.8. For more uncertain parallaxes, we find evidence that distances (parallaxes) are generally overestimated (underestimated). Furthermore, we find strong evidence that the large error inflation factors of AGB stars reported in the literature are unrealistic. Our results lend confidence to the \Gaia parallax measurements of these highly extended variable stars.}

\authorrunning{Uttenthaler et al.}
\titlerunning{IR period--luminosity relations of Galactic Miras}
\maketitle

\section{Introduction}\label{sec:intro}

Long-period variables (LPVs) are found on the upper part of the giant branch, particularly the asymptotic giant branch (AGB), and thus represent a late evolutionary stage of low- and intermediate-mass stars ($1-8M_{\odot}$). Their variability is caused by pulsational instabilities in their cool and extended outer envelopes. Pulsation proceeds in one or more radial or non-radial modes, and it is thus expected that LPVs follow a set of logarithmic period--luminosity (PL) relations in respective diagrams \citep{2000PASA...17...18W}. Long-time monitoring of stellar systems such as the Magellanic Clouds (MCs) has opened the possibility of deriving these relations observationally \citep[e.g.][]{2004AcA....54..129S,2010ApJ...723.1195R,2013ApJ...763..103S} because they host large LPV populations at nearly identical distances with relatively small and uniform interstellar extinction. Comparisons with pulsation models \citep[e.g.][]{2017ApJ...847..139T} enable us to gain a deep understanding of the fundamental aspects of AGB stars, including their stellar parameters, atmospheric structure, and mass loss.

Among the LPVs, the Mira variable class is of particular importance because these stars can be easily distinguished from other variables by their very large light amplitudes in the visual range combined with long pulsation periods of 100 to 1000 days. Thus, Miras in the solar neighbourhood can be monitored even with amateur means \citep{2013JAVSO..41..348K}. There is a general consensus today that Miras pulsate in the radial fundamental mode \citep[FM;][]{2017ApJ...847..139T} and form one of the prominent PL relations.

In the past, the study of stellar systems such as the MCs was most fruitful for understanding PL relations of LPVs. As a result of identical distances and similar reddening, the PL relations are very well defined, and individual stars can be uniquely assigned to the respective relations (that is, pulsation mode) once their periods are known. Homogeneous long-term surveys have provided periods and average magnitudes for LPVs in great numbers \citep[e.g.][]{2003AcA....53..291U}. In contrast, Galactic LPVs have received much less attention because their distances were poorly known.

To measure the distances to LPVs in the solar neighbourhood, both ground- and space-based approaches have been applied in the past decades. From the ground, very long baseline interferometry (VLBI) was used to measure the parallaxes (e.g. \citet{2016PASJ...68...78N} and \citet{2020PASJ...72...50V}). \citet{2022A&A...667A..74A} established a bolometric PL relation of Galactic Miras based on these VLBI parallaxes, but due to the extensive observational effort of this method and the limitation in the number of accessible objects, sample sizes are very small in these studies. Space-based parallaxes for LPVs first became available with the Hipparcos mission. Again, the sample remained small and limited to the nearest objects, and attempts to derive a PL relation solely based on data for Galactic field stars remained inconclusive \citep[e.g.][]{2007MNRAS.378.1543G}. Nevertheless, by adopting the PL slope from observations of the MCs, \citet{2000MNRAS.319..759W} and \citet{2010MNRAS.409..777T} successfully determined the Large Magellanic Cloud (LMC) distance modulus from the PL zero-point of Galactic Mira-like and semi-regular variables (SRVs), respectively, from their Hipparcos parallaxes.

The study of Galactic LPVs was recently revived by the advent of large catalogues of parallax measurements provided by the \Gaia space observatory \citep{2023A&A...674A...1G}. \citet[][see their Fig.~32]{2023A&A...674A..15L} showed that the PL relations of Galactic LPVs can be revealed when using the \Gaia parallax measurements. Based on their \Gaia catalogue and the 2MASS catalogue \citep{2006AJ....131.1163S}, \citet{2023MNRAS.523.2369S} applied a probabilistic model to infer the Galactic Mira PL relation in the 2MASS $K$ band. This allowed many insights to be gained and even the application of the Mira PL relation to measure the local value of the Hubble constant \citep{2024ApJ...963...83H}. Such applications underline the importance of Mira PL studies much beyond the immediate field of the late phases of low-mass stellar evolution.

However, three important issues need to be mentioned that hamper observational derivation of a reference PL relation for Miras to be used for comparison with models and distance determination. First, Miras are strongly variable, even in near-IR bands. Many studies largely rely on the single-epoch 2MASS photometry due to the challenge of observing a star's brightness variation over more than a year. However, the use of single-epoch measurements leads to a significant broadening of the observed PL relations by variability scatter. Second, the LPV class is not as homogeneous in mass as the Cepheids. Mass scatter widens the PL relation. Third, LPVs on the AGB are expected to have large parallax uncertainties due to their intrinsic properties, which in turn include multiple factors. Surface brightness variations induced by large convective cells may significantly impact the measurement of the stellar photocentre and create noise in the \Gaia astrometry \citep{2018A&A...617L...1C} since the radius of these stars is of the order of one astronomical unit. The stars are also intrinsically very red, and \Gaia's imaging system is not well characterised for such red sources. Finally, Miras vary in colour during a pulsation cycle \citep{2019gaia.confE..62M}, but the current DR3 astrometric pipeline uses a fixed colour. These measurement-related problems have an unclear effect on the \Gaia parallax uncertainty, though the sophisticated estimates that have been made \citep{2023MNRAS.523.2369S}. \citet{2022A&A...667A..74A} suggest that the uncertainties are underestimated by up to a factor of 5.44 for Miras brighter than $G=8\fm0$ on average. The crucial question of potentially underestimated \Gaia parallaxes is actively discussed in the community \citep{2018evn..confE..43V,2021MNRAS.506.2269E,2022A&A...657A.130M,2023MNRAS.523.2369S}.

In the present paper, our aim is to provide PL relations for nine near-IR bands, some for the first time, and to address the issue of photometric variability by using multi-epoch IR photometry of Galactic Miras from two catalogues. We combined this photometry with period determinations from simultaneous optical observations of these Miras. By largely eliminating the variability-induced scatter of stars around the PL relation, our approach allowed us to address the issue of underestimated parallax uncertainties with special regard to Miras.

\section{Data}\label{sec:data}

\subsection{Selection of the Mira sample}

The first IR catalogue we used is that published by \citet{2010ApJS..190..203P}. They extracted weekly averaged fluxes for 2652 stars from the cold and warm era all-sky maps of the Diffuse Infrared Background Experiment (DIRBE) on board the Cosmic Microwave Background Explorer (COBE) satellite in four near-IR bands (1.25, 2.2, 3.5, and $4.9\,\mu{\rm m}$). The cold and warm DIRBE eras span 3.6 years of observations, which is long enough to cover more than one cycle of even the longest-period FM pulsators (Miras). We adopted their list of variable and candidate variable stars from their Tables~1 and 2, comprising 597 objects in total. We call this the DIRBE catalogue in the following. According to \citet{2004ApJS..154..673S}, DIRBE photometry of the brightest IR sources is ten times more precise than that of 2MASS, and the relations derived by the same authors suggest that the differences between the DIRBE and 2MASS filter systems are small enough to be neglected for our purposes. {An inspection of the $[2.2]-K$ versus $[2.2]$ and $[1.25]-J$ versus $[1.25]$ diagrams of the DIRBE non-variable stars \citep[][their Table~3]{2010ApJS..190..203P} confirms that the mean magnitude differences and the trends with magnitude are statistically indistinguishable from zero, in particular for bright stars with $K<1\fm0$.}

The unTimely catalogue presented by \citet{2023AJ....165...36M} serves as the second source of IR photometry. It is a deep time-domain catalogue of detections based on Wide-field Infrared Survey Explorer \citep[WISE;][]{2010AJ....140.1868W} and NEOWISE observations in the $W1$ ($3.4\,\mu{\rm m}$) and $W2$ ($4.6\,\mu{\rm m}$) bands spanning $\sim10.5$ years from 2010 through 2020. Most of the sample stars have been observed 16 times, with two observations in 2010 and two per year in the period 2014 to 2020, separated by about half a year. Assuming an approximately sinusoidal light curve, our numerical tests showed that the true mean brightness is much more precisely reconstructed if it is calculated from the mean of the maximum and minimum flux, rather than from the mean of all $\sim16$ observations of each star. The reason is that the sine wave is relatively flat around the extrema, but a quasi-random sampling at epochs away from the extrema, where the flux changes much more quickly with time, introduces significant uncertainty in the mean brightness. The average magnitudes of the stars were thus converted from the average of the maximum and minimum flux as given in the instructions in the unTimely catalogue based on the definition of nanomaggies \citep{2004AJ....128.2577F}.

\begin{figure}
\centering
\includegraphics[width=\linewidth]{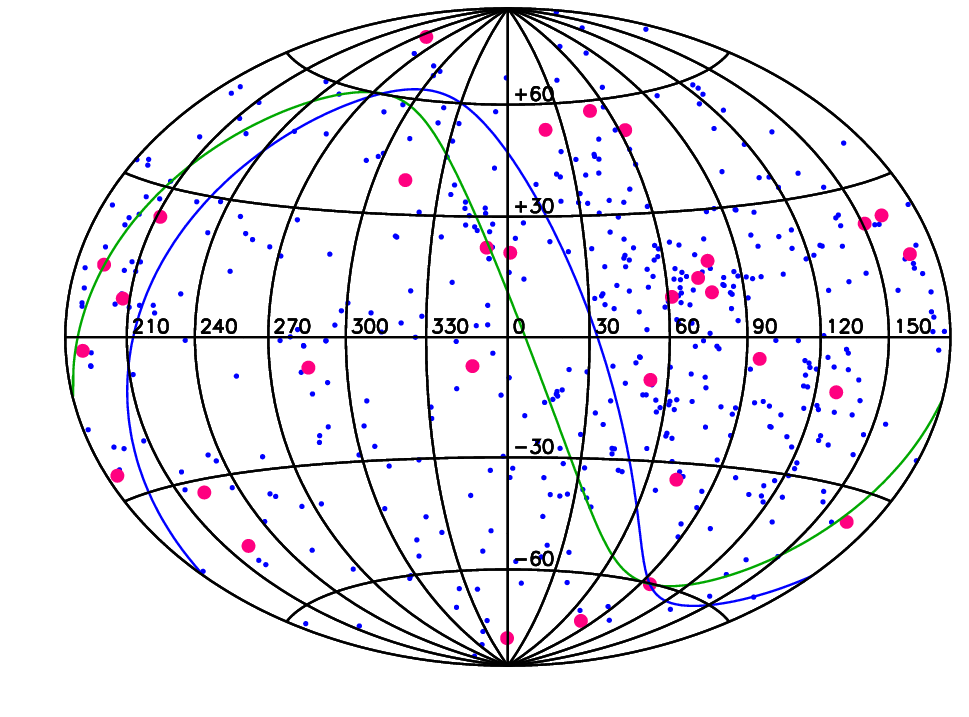}
\caption{Aitoff projection in Galactic coordinates of the Miras (filled circles) in the compilation by \citet{2013JAVSO..41..348K}. The large pink symbols represent stars within $\pm12$\,d of a one-year period. The solid green line is the ecliptic, and the solid blue line is the celestial equator.}
\label{fig:Aitoff}
\end{figure}

Variability periods have been determined and listed by \citet{2010ApJS..190..203P} for the DIRBE catalogue. However, for the unTimely sources, the cadence of observations is too low to determine reliable periods from the IR observations themselves, except for the longest-period stars. Therefore, we resorted to a compilation of Mira periods obtained from optical observations taken contemporaneously with the IR data. We used the tables of the period evolution of 510 solar neighbourhood Miras curated by T.\ Karlsson\footnote{\url{https://var.saaf.se/mirainfoper.php}}. The data collection is described in more detail in \citet{2013JAVSO..41..348K}. The selection of stars is based on available observations in the AAVSO\footnote{\url{https://www.aavso.org/}} database. All variables were discovered several decades ago and classified by those classical observers according to the amplitude criterion for Miras of $\Delta V\ge2\fm5$ \citep{2017ARep...61...80S}. Stars must have 20 clear maxima documented to be included in the selection. These selection criteria may impose some biases. For historical reasons, more stars in the northern than in the southern hemisphere are included (325/510=63.7\% versus 185/510=36.3\%). Figure~\ref{fig:Aitoff} shows an Aitoff projection of the complete sample in Galactic coordinates, which demonstrates this distribution with respect to the celestial equator (solid blue line). Furthermore, stars close to the ecliptic and objects with periods around one year may be under-represented in our sample: the former due to the longer annual observing gaps and the latter in cases where the maximum falls within these gaps. Indeed, the period histogram of the sample stars (not shown) indicates a trough at periods slightly shorter than one year, suggesting that these objects are under-represented in our sample. However, from Fig.~\ref{fig:Aitoff}, we cannot identify a `zone of avoidance' around the ecliptic (solid green line), in particular not for Miras with periods close to one year (large pink symbols).

Concerning selection effects, we stress that our sample is limited to optically bright Mira variables. This is a consequence of our need to minimise the effects of photometric variability in our Galactic PL relations, which requires datasets only available for an optically bright sample of stars. Thus, stars with high mass-loss rates and strong circumstellar dust obscuration are naturally missing from this collection. In this sense, our sample is not representative of all FM variables. Nevertheless, this selection strategy is favourable for our purpose, as we search for relations between period and brightness in individual photometric bands in the near-IR to be used as representatives of the PL relation. This will only work if the SED is not significantly affected by circumstellar dust. In summary, we considered the potential biases in the sample to be either relatively minor issues that are hard to avoid in any sample of solar neighbourhood Miras with well-determined periods or favourable for our purposes.

We found 132 stars in the DIRBE catalogue with periods listed in the Karlsson compilation. Conversely, we downloaded the WISE photometry for the 510 Karlsson Miras using the Python unTimely Catalog Explorer. We adopted the pulsation period with the Julian date closest to the mean date of the respective IR observations.

\begin{figure}
\centering
\includegraphics[width=\linewidth]{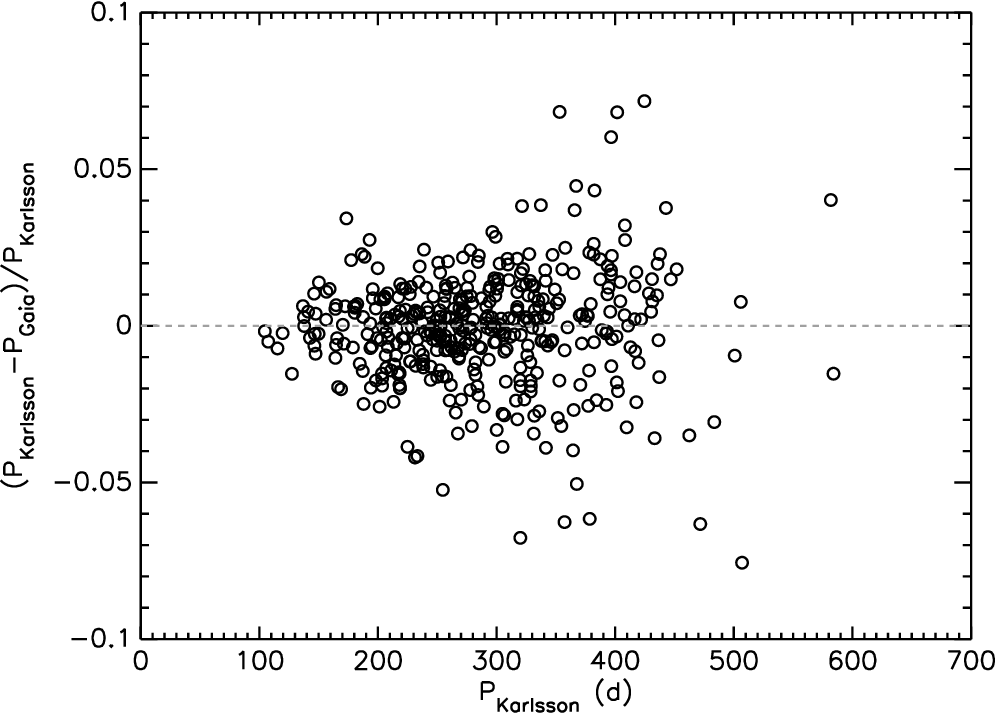}
\caption{Comparison of the periods in the Karlsson compilation with those of the \Gaia catalogue of LPV candidates by \citet{2023A&A...674A..15L}. The diagram plots the relative difference between both periods as a function of the period as tabulated by \citet{2013JAVSO..41..348K}.}
\label{fig:Gaia-LPV}
\end{figure}

We also compared the periods of the Karlsson compilation with those reported by the \Gaia catalogue of LPV candidates \citep{2023A&A...674A..15L}. Among the 510 stars, 13 had not been detected as LPV candidates by the Gaia data analysis. A further 72 stars had been identified as LPVs in the Gaia classification table, but did not make it into the final Gaia LPV catalogue due to the additional selection criteria applied. A discussion on some prominent LPVs missing from the LPV catalogue can be found in \citet{2023A&A...674A..15L}. From the remaining 425 stars, 422 have a published Gaia period and can thus be used for our comparison. The periods of two stars, R~Cen and S~Sco, differ by a factor of approximately two between the two databases, where \citet{2023A&A...674A..15L} lists the shorter period. R~Cen has a double-peaked light curve, so that the \Gaia pipeline probably picked up half of the true period of $\sim500$\,d. In Fig.~\ref{fig:Gaia-LPV}, we show a comparison between the periods tabulated in both databases. In general, the agreement is very good. The relative difference has a root mean square (RMS) scatter of $\sim1.8$\%. However, we also see that the scatter clearly increases for periods $\gtrsim300$\,d. For some stars, the periods differ by as much as 7\%. This increased scatter is most probably also caused by the limited time baseline of the \Gaia observations considered in \citet{2023A&A...674A..15L}. This conclusion is confirmed by inspecting the period uncertainties quoted in the \Gaia catalogue of LPV candidates (not shown in Fig.~\ref{fig:Gaia-LPV} for clarity).

\subsection{Preparation of the data}

Both the DIRBE and the unTimely sample were cross-matched with the catalogue of \Gaia distances and their $1\sigma$ uncertainties by \citet{2021AJ....161..147B}, as well as the 2MASS catalogue \citep{2006AJ....131.1163S}. The distances and their uncertainties reported by \citet{2021AJ....161..147B} were determined from \Gaia EDR3 parallaxes and their uncertainties\footnote{It has been shown that \Gaia parallaxes are affected by a zero-point bias and that their uncertainties are tentatively underestimated \citep{2021A&A...649A...5F}. Most of these issues have been taken into account by \citet{2021AJ....161..147B}.} using Bayesian statistics, with priors constructed from a mock catalogue based on the Besan\c{c}on Galactic model. {The quoted asymmetric uncertainties are the 16th and 84th percentiles of the posterior distance distribution. Because the distance modulus is a monotonic transformation of distance, these percentiles map directly to the corresponding percentiles in distance modulus.} Magnitudes in the DIRBE bands were calculated using the zero-magnitude fluxes given by \citet{2010ApJS..190..203P}.

The photometry was de-reddened using the 3D map of \citet{2017AstL...43..472G}. As this map extends only to 1200\,pc from the Sun, the extinction was set to the outer boundary for the few more distant stars. We applied the extinction law of \citet{2016ApJS..224...23X} to the 2MASS and WISE bands. As no relative extinction values have been determined specifically for the DIRBE bands, we adopted the same relative extinction values for the $[1.25]$ and $[2.2]$ bands as for the 2MASS $J$ and $K$ bands, and we interpolated the extinction values to $A_{[3.5]}/A_{Ks}=0.60$ and $A_{[4.9]}/A_{Ks}=0.48$ from Fig.~18 of \citet{2016ApJS..224...23X} for the DIRBE $[3.5]$ and $[4.9]$ bands, respectively. The median extinction in the $[2.2]$ band for the DIRBE gold sample (see below) is $0\fm044$. Figure~\ref{fig:Aitoff} shows that there is a `zone of avoidance' around the Galactic equator in our sample, and most stars are located at $|b|=15-20^\circ$; thus, strongly extinguished stars are missing in our sample. Although there are still significant variations in absolute extinction between the different maps available in the literature \citep{2023AstL...49..673G}, \citet{2025A&A...700A.187U} found negligible differences between the map of \citet{2017AstL...43..472G} and the updated map of \citet{2023AstL...49..673G} in near-IR extinction corrections for Galactic Miras. Therefore, we are confident that any other choice of extinction map or relative extinction values would have very modest effects on our results.

We defined a gold sample in all bands by setting a limit of 10\% uncertainty on the \Gaia-based distance measurement and by requiring a period available from the Karlsson compilation. We discarded one more star from the DIRBE sample, namely R~For, which appears to be a faint outlier by more than $3\sigma$ in the DIRBE [2.2] band. This star was identified by \citet{1984MNRAS.211..331F} as undergoing changes in circumstellar dust obscuration. An inspection of the AAVSO $V$ band light curve shows that R~For was in a relatively faint phase from 1990 to 1995, exactly when DIRBE observed it. Nevertheless, we did retain it for the derivation of the 2MASS PL relations, as the 2MASS observations were taken some ten years later, when the star was at its normal brightness again. This procedure yielded a gold sample of 110 stars with DIRBE observations {\citep[91 M\mbox{-}, nine MS/S-, and ten C-type; spectral types from][]{2014yCat....1.2023S}} and 312 stars with unTimely observations (253 M-, 28 MS/S\mbox{-}, 30 C-type, and one unknown spectral type). Additional constraints on available and reasonable 2MASS photometric uncertainties removed a few stars from the 2MASS gold sample, retaining 308 stars (253 M-, 28 MS/S-, 27 C-type). The DIRBE gold sample stars are at distances between 150 and 1550\,pc, with a mean of 710\,pc and a standard deviation of 340\,pc, while the unTimely and 2MASS gold sample stars range between 150 and 3260\,pc, with a mean of 1060\,pc and a standard deviation of 560\,pc. Thus, the unTimely gold sample is, on average, more distant than the DIRBE gold sample.

The uncertainties on the absolute magnitudes of the various DIRBE and WISE bands were determined by combining in quadrature the uncertainty induced by the distance measurement and the photometric uncertainty, where for the latter, we adopted the standard deviation of the mean of all observations of a star in a given band. In both the DIRBE and the unTimely samples, the uncertainty from the distance measurement is the dominating source of uncertainty on the absolute magnitude, being on average a factor of 4.6 and 2.4 larger than the photometric uncertainty in the two samples, respectively. In absolute terms, the distance uncertainty implies an average uncertainty on the absolute magnitude of $0\fm117$ in the DIRBE sample and of $0\fm133$ in the unTimely sample. We also inspected the renormalised unit weight error (\texttt{RUWE}) of our sample stars. The unTimely gold sample stars have a median \texttt{RUWE} of 1.1, and 76.8\% have \texttt{RUWE}$<$1.4. A fraction of 92.9\% have \texttt{RUWE}$<$2.0 -- a limit still likely safe to use according to \citet{2021A&A...649A..13M}.

We had to further restrict the gold sample in the $W2$ band because, despite the efforts in the unTimely catalogue to reconstruct the total flux of bright, partially saturated sources, it suffers from additional saturation effects. By plotting the magnitude differences (colours) of the 2MASS $K_{\rm S}$, $W1$, and $W2$ bands versus their magnitudes {for our sample Miras as well as for the bright non-variable stars of \citet{2010ApJS..190..203P}}, we found that while $K_{\rm S}-W1$ has no obvious dependence on apparent magnitude, colours involving the $W2$ band showed a significant trend for sources brighter than $W2=1\fm4$. Thus, we excluded sources from the $W2$ gold sample brighter than this obvious saturation limit. This left us with 152 stars in the $W2$ gold sample (129 M-, 14 MS/S-, and nine C-type stars).

This procedure ensured that the most obvious saturation effects were addressed. As has been shown by \citet[][their Fig.~7]{2019ApJS..240...30S}, saturation sets in at $W1=8\fm0$, and all our sample stars are brighter than this limit. In Sect.~\ref{sec:synthSEDs}, we make additional checks if the $W1$ and $W2$ fluxes are reliable in absolute terms. As we also show below (e.g. Fig.~\ref{fig:unTimely_PLW1}), the WISE fluxes should be reliable at least in a relative sense among the stars.

\section{Results}\label{sec:results}

\subsection{Period--luminosity diagrams}\label{sec:PLDs}

\begin{figure}
\centering
\includegraphics[width=\linewidth]{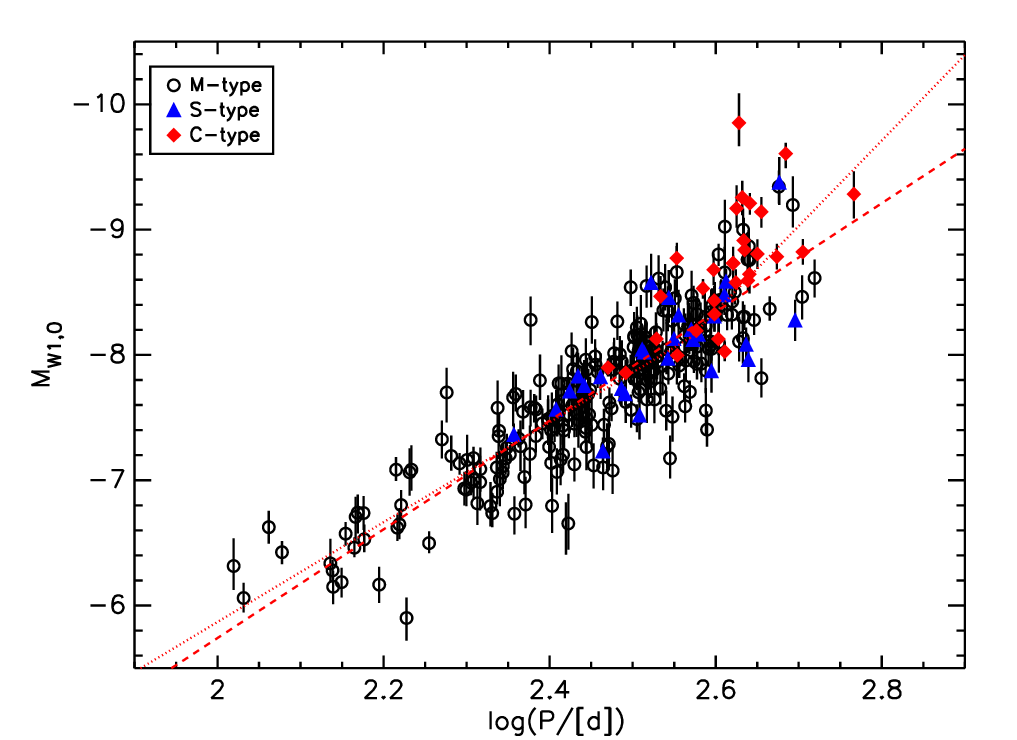}
\caption{Period--luminosity diagram using the de-reddened, absolute magnitude in the WISE $W1$ band, $M_{{\rm W1},0}$. Stars of different chemical spectral types are represented by different colours and symbol shapes (see the legend). The plotted error bars include the uncertainty of the mean photometric magnitude and the (asymmetric) uncertainty due to the distance of the stars. The dashed red line is the weighted linear least-square fit reported in Table~\ref{tab:PL relations}. The dotted red line is a fit by two independent linear relations, indicating that the slope increases at $\log(P)\gtrsim2.57$; see Sect.~\ref{sec:PLR}.}
\label{fig:unTimely_PLW1}
\end{figure}

\begin{figure}
\centering
\includegraphics[width=\linewidth]{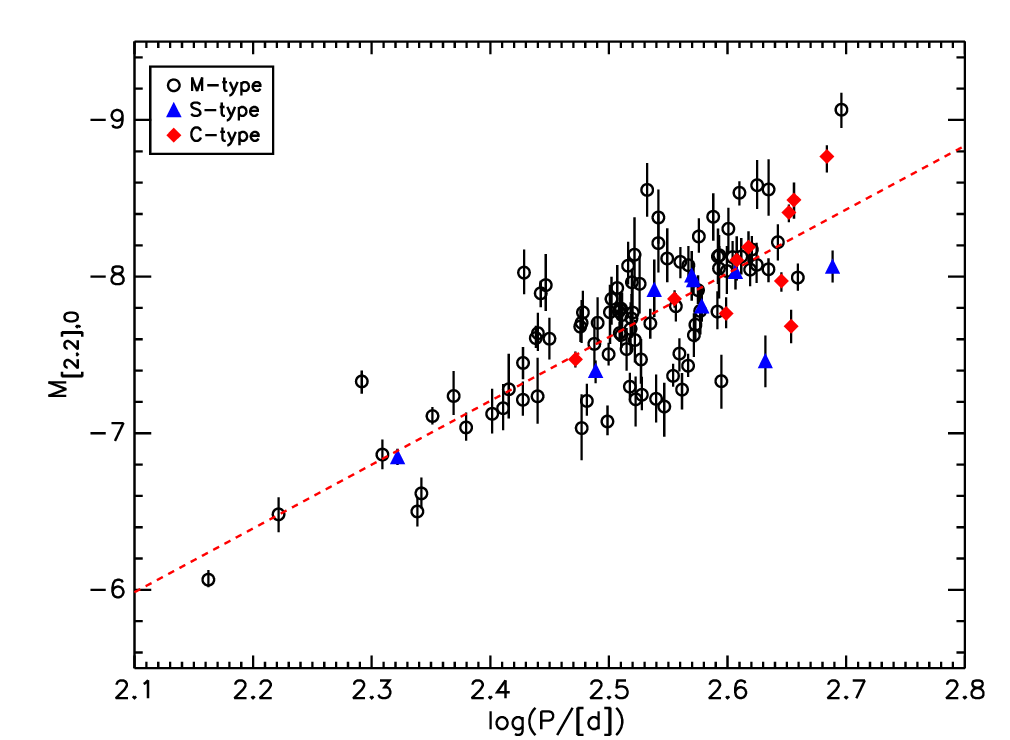}
\caption{Same as Fig.~\ref{fig:unTimely_PLW1} but for the DIRBE [2.2] band, $M_{[2.2],0}$.}
\label{fig:DIRBE_PL2.2}
\end{figure}

Our main results are shown in Figs.~\ref{fig:unTimely_PLW1} and \ref{fig:DIRBE_PL2.2}. They show the de-reddened, absolute magnitudes in the WISE $W1$ band ($M_{{\rm W1},0}$) and DIRBE [2.2] band ($M_{[2.2],0}$), respectively. Both samples exhibit well-defined sequences of brighter absolute magnitude with increasing pulsation period, as is known from many other works.

These are probably the best-defined PL relations of substantial samples of Galactic Miras ever presented. The stars scatter around the least-squares linear fit (dashed red line; see Sect.~\ref{sec:PLR}) in the  WISE $W1$ band with a standard deviation of $0\fm327$ and $0\fm313$ in the DIRBE [2.2] band, respectively. The (single-epoch) 2MASS sample stars scatter with a significantly increased standard deviation of $0\fm403$ in the $K_{\rm S,0}$ band, which corresponds to an additional scatter with a standard deviation of $0\fm254$ added in quadrature. This additional scatter corresponds to the standard deviation of a sinusoidal curve with a full amplitude of $0\fm254\times2\times\sqrt{2}=0\fm718$, which should be a good measure of the average amplitude of the stars in the $K_{\rm S}$ band. Applying sinusoidal fits to the individual DIRBE [2.2] light curves, we find that the mean amplitude is $0\fm72$ and the median amplitude is $0\fm64$, in excellent agreement with the estimate from the scatter.

We experimented with the $W1$ photometry in a similar manner. To simulate the scatter of single-epoch $W1$ observations, we extracted only the last such observation from the tables and re-ran the linear least-squares fit on these single-epoch observations. Here, the scatter increased from $0\fm327$ to $0\fm401$, corresponding to an additional scatter with a standard deviation of $0\fm234$ added in quadrature. This corresponds to the standard deviation of a sinusoidal curve with a full amplitude of $0\fm66$, which should again be representative of the average amplitude of the gold sample stars in the $W1$ band. Generally, Mira amplitudes decrease with increasing wavelength, but the amplitude difference between the $K$ and $W1$ bands is not significant \citep{2021ApJS..257...23I}.

\subsection{Period--luminosity relations}\label{sec:PLR}

We determined for the gold samples the PL relations in the DIRBE [1.25], [2.2], [3.5], and [4.9], the $W1$ and $W2$, as well as the 2MASS $J$, $H$, and $K$ bands by weighted linear last-squares fits, using the \texttt{SIXLIN.pro} GDL routine. Assuming Gaussian error distributions, we weighted the individual stars by $1/\sigma^2$, where $\sigma$ is the uncertainty on the absolute magnitude in the respective band. As the uncertainty stemming from the distance uncertainty is asymmetric, we symmetrised it for calculating the weight. We did not consider period uncertainties, as this is expected to be a minor effect. Judging from the period evolution of the stars (footnote~1), their uncertainty is of the order of $\lesssim1$\%.

We chose to compare the results with those of \citet{2008MNRAS.386..313W}, who also analysed a sample of Galactic Miras. Thus, we write the PL relations in the form
\begin{equation}\label{Eq:PL}
    M_{0} = b\times(\log P - 2.38) + a,
\end{equation}
where $b$ is the slope of the relation and $a$ is the zero-point at $\log(P)=2.38$. Other works applied quadratic rather than linear fits to the data, for example, \citet{2017AJ....153..170Y}. A departure from a linear trend is not immediately obvious in our PL diagrams, most probably because our sample lacks very long-period Miras that are present in other works. Nevertheless, we made a fit with two independent linear relations of the PL($W1$) diagram (Fig.~\ref{fig:unTimely_PLW1}). Based on the results of \citet{2023MNRAS.523.2369S}, initial values for a simplex downhill minimisation routine were chosen such that a break in the slope occurs near $\log(P)=2.6$. Indeed, the fit resulted in a steeper slope of $-6.82\pm0.05$ for the stars with periods longer than $\log(P)=2.57$. The relations for both regimes, $\log(P)<2.57$ and $\log(P)\ge2.57$, are plotted as dotted lines in Fig.~\ref{fig:unTimely_PLW1}. Except for the DIRBE Wesenheit index (see below), we did not make separate fits to the S and C stars because they are too few and follow the trend of the M-type stars.

The results of our weighted least-squares fits are collected in Table~\ref{tab:PL relations}. The quoted uncertainties are the standard deviations as determined by the \texttt{SIXLIN.pro} routine. The relations are sorted by increasing effective wavelengths of the respective photometric band. From this sorting, we see that, generally, the slope of the relation $b$ becomes steeper and the zero point $a$ becomes brighter with increasing wavelength. This result is broadly consistent with those reported by \citet{2010ApJ...723.1195R}, \citet{2017AJ....153..170Y}, and \citet[][and references therein]{2021ApJS..257...23I}. In addition, the slopes and the zero points for the DIRBE [1.25] and the 2MASS $J$, as well as for the DIRBE [2.2] and the 2MASS $K_{\rm S}$ bands, agree very well. The period range for the DIRBE relations is $2.16\lesssim\log(P)\lesssim2.70$, whereas for the unTimely and 2MASS relations it is $2.02\lesssim\log(P)\lesssim2.77$.

\begin{table}
\caption{Period--luminosity relations of the form $M_{0} = b\times(\log P - 2.38) + a$ (Eq.~\ref{Eq:PL}) of Galactic Miras, sorted by increasing effective wavelength of the NIR band.}
\label{tab:PL relations}
\centering
\begin{tabular}{llcc}
\hline\hline
Band              & $\lambda_{\rm eff}$ & $b$ & $a$ \\
                  & ($\mu{\rm m}$)      &     &     \\
\hline
2MASS $J$         & 1.235 & $-2.690\pm0.035$ & $-5.801\pm0.006$ \\
DIRBE [1.25]      & 1.25  & $-2.748\pm0.037$ & $-5.860\pm0.005$ \\
2MASS $H$         & 1.662 & $-3.549\pm0.036$ & $-6.696\pm0.006$ \\
2MASS $K_{\rm S}$  & 2.159 & $-4.010\pm0.032$ & $-7.144\pm0.005$ \\
DIRBE [2.2]       & 2.2   & $-4.072\pm0.022$ & $-7.125\pm0.004$ \\
unTimely $W1$     & 3.368 & $-4.337\pm0.017$ & $-7.389\pm0.002$ \\
DIRBE [3.5]       & 3.5   & $-4.827\pm0.019$ & $-7.494\pm0.003$ \\
unTimely $W2$     & 4.618 & $-3.977\pm0.029$ & $-8.124\pm0.004$ \\
DIRBE [4.9]       & 4.9   & $-5.124\pm0.030$ & $-7.702\pm0.005$ \\
\hline
\end{tabular}
\end{table}

\begin{table}
\caption{Period--luminosity relations from the literature, converted to the form $M_{0} = b\times(\log P - 2.38) + a$ (Eq.~\ref{Eq:PL}).}
\label{tab:PLR-lit}
\centering
\begin{tabular}{lcccc}
\hline\hline
Ref.\ & Band        & $b$              & $a$              & System \\
\hline
W08   & $K$         & $-3.51\pm0.20$   & $-7.25 \pm0.07 $ & Galaxy \\
R10   & $K_{\rm S}$ & $-3.31\pm0.04$   & $-7.485\pm0.09 $ & LMC    \\
I21   & $W1$        & $-3.807\pm0.066$ & $-7.554\pm0.025$ & LMC    \\
I21   & $W2$        & $-3.794\pm0.096$ & $-7.751\pm0.027$ & LMC    \\
\hline
\end{tabular}
\tablefoot{References: W08: \citet{2008MNRAS.386..313W}; R10: \citet[][their Sequence 1 for O-rich stars]{2010ApJ...723.1195R}; I21: \citet[][their linear fit to the O-rich stars with $\log(P)\leq2.6$]{2021ApJ...919...99I}.}
\end{table}

We compared our results with those for the Galaxy and the LMC reported in the literature and collected in Table~\ref{tab:PLR-lit}. As our samples are dominated by O-rich Miras, we compared with literature results for O-rich Miras. The zero point of the PL relation in the DIRBE $[2.2]$ and $K_{\rm S}$ bands agrees very well with the one found by \citet{2008MNRAS.386..313W} for O-rich Galactic Miras within $\sim0\fm1$. We find a steeper slope than \citet{2008MNRAS.386..313W}, $b\simeq-4.0$ versus $b\simeq-3.5$, but their slope was fixed from a sample of LMC Miras. In fact, the slope reported by \citet[][their Sequence 1]{2010ApJ...723.1195R} for a sample of 2218 O-rich LMC Miras ($b=-3.31$) is in better agreement with the slope found by \citet{2008MNRAS.386..313W}. Assuming a distance modulus to the LMC of $\mu=18\fm477$ \citep{2019Natur.567..200P}, the $K_{\rm S}$ zero point of the LMC Miras of \citet{2010ApJ...723.1195R} is brighter than that of the Galactic Miras by $\sim0\fm235-0\fm360$. \citet{2023MNRAS.523.2369S} estimate that Galactic Miras are fainter than LMC Miras in the $K_{\rm S}$ band by $0\fm11$ at $P=250$\,d ($\log P=2.4$). We thus confirm the conclusion by \citet{2023MNRAS.523.2369S} that the near-IR PL relations of O-rich Galactic Miras at short periods are steeper than those in the LMC and that Galactic Miras are fainter in these bands, suggesting that metallicity or population effects play a role. Recent FM pulsation models by \citet{2025AstL...51..177F} calculated for solar and LMC metallicity indicate a metallicity effect pointing in the same direction.

Table~\ref{tab:PLR-lit} also contains the $W1$ and $W2$ relations of LMC Miras reported by \citet{{2021ApJ...919...99I}}. The zero point of the Galactic Miras in the $W1$ band is fainter than that of the LMC Miras by $0\fm165$. However, as we show in Sect.~\ref{sec:synthSEDs} below, the $W1$ band may suffer from additional saturation effects, and we must take this result with a grain of salt. Taking the results for the $W2$ band at face value, the Galactic Miras would be brighter than the LMC Miras of \citet{2021ApJ...919...99I} by $0\fm37$. However, the $W2$ zero point does not comply well with the trend of monotonically increasing brightness with increasing wavelength visible in Table~\ref{tab:PL relations}. As noted above, the $W2$ band suffers from saturation, but we took this into account by eliminating stars with $W2<1\fm4$. The inspection of light curves of stars close to this saturation limit did not reveal any additional issues, and the $W1-W2$ versus $M_{W1,0}$ colour-magnitude diagram looks inconspicuous. As we show in Sect.~\ref{sec:synthSEDs}, the $W2$ flux is significantly brighter than a Planck curve fit to the other data points. We conclude that the $W2$ flux of our sample stars could be significantly overestimated. We include the PL($W2$) relation in Table~\ref{tab:PLR-lit} only for completeness and urge the reader to use it with caution.

It may seem that the scatter in the PL diagram of Galactic Miras (e.g. Fig.~\ref{fig:unTimely_PLW1}) is larger than that of LMC Miras. For example, \citet{2010ApJ...723.1195R} report residual scatter of $0\fm272$ of the O-rich stars around their (Mira) Sequence~1 in the 2MASS $K_{\rm S}$ band. This may be compared to the RMS scatter of $0\fm403$ of our 2MASS gold sample (the numbers are very similar for our WISE $W1$ and their Spitzer [3.6] band). Subtracting in quadrature the mean parallax-induced uncertainty on the absolute magnitude reduces the scatter to $0\fm380$, still significantly larger than the scatter reported by \citet{2010ApJ...723.1195R}. However, one cannot straightforwardly attribute this to population effects because \citet{2010ApJ...723.1195R} and other works on LMC samples introduce hard boundaries to their sequences before fitting them, effectively applying a sigma-clipping. Therefore, it is difficult to make direct comparisons at this point.

\begin{figure}
\centering
\includegraphics[width=\linewidth]{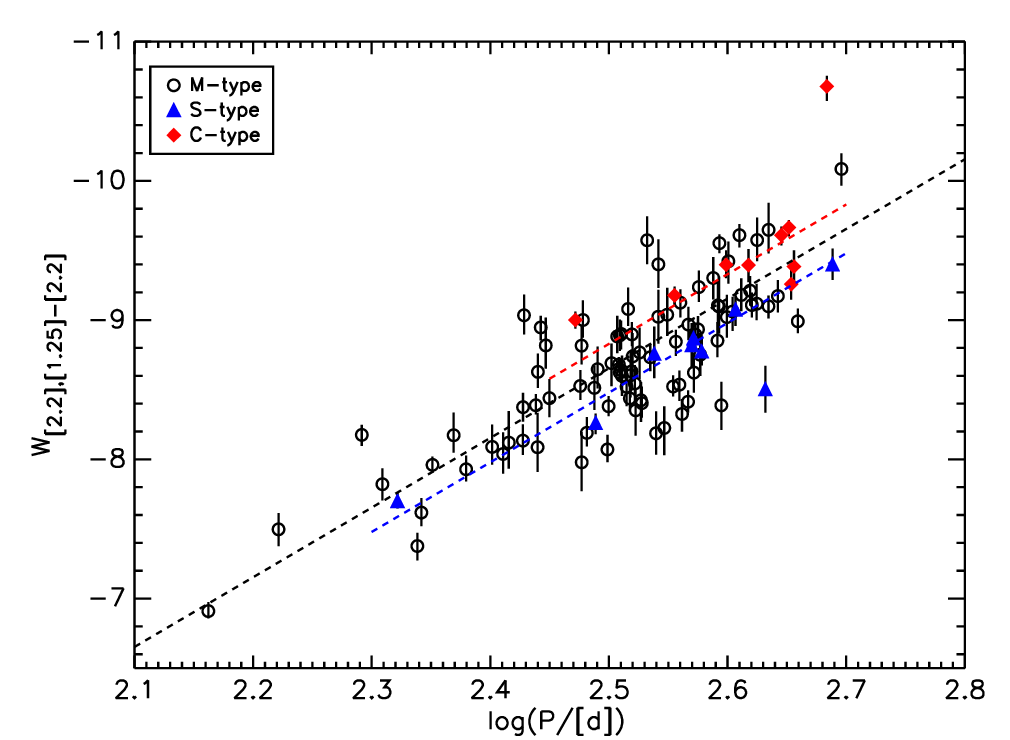}
\caption{DIRBE Wesenheit index $W_{[2.2],[1.25]-[2.2]}$ vs $\log P$ of the DIRBE gold sample. Stars of different chemical spectral types are represented by different colours and symbol shapes (see the legend). The dashed black, blue, and red lines show linear fits to the overall DIRBE gold sample, the S stars, and the C stars, respectively (see Table~\ref{tab:Wesenheit}). The same slope as for the overall sample was adopted for the S and C stars.}
\label{fig:DIRBE_Wesenheit}
\end{figure}

\begin{figure}
\centering
\includegraphics[width=\linewidth]{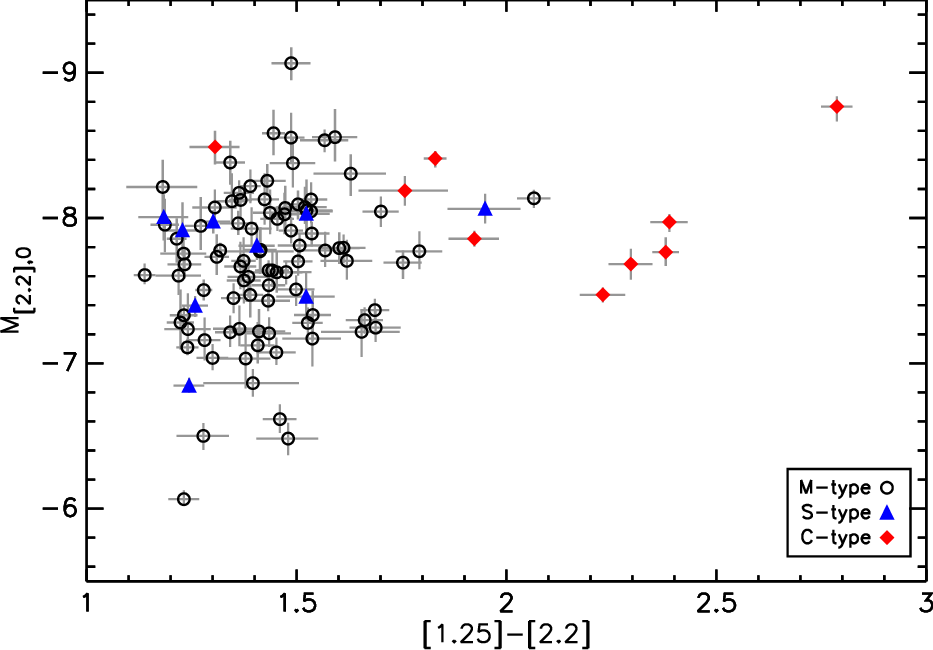}
\caption{DIRBE $[2.2]$ vs $[1.25]-[2.2]$ colour-magnitude diagram of the DIRBE gold sample. Symbols are as in Fig.\ref{fig:DIRBE_Wesenheit}.}
\label{fig:DIRBE_CMD}
\end{figure}

Finally, we constructed a Wesenheit index based on the DIRBE [1.25] and [2.2] photometry. It is defined as
\begin{equation}\label{Eq:Wesenheit}
W_{[2.2],[1.25]-[2.2]}=[2.2]-0.686\times([1.25]-[2.2]),
\end{equation}
following \citet{2005AcA....55..331S}, who obtained the near-IR Wesenheit index based on the \citet{1998ApJ...500..525S} Galactic extinction law. The Wesenheit index eliminates the interstellar reddening component and a significant portion of the circumstellar reddening component, provided that the two extinction laws are not too dissimilar \citep{2018A&A...616L..13L}. Figure~\ref{fig:DIRBE_Wesenheit} shows $W_{[2.2],[1.25]-[2.2]}$ of the DIRBE gold sample as a function of the pulsation period. The magnitude uncertainties are very similar to those for $M_{[2.2],0}$. It is evident from this diagram that most of the C stars are brighter by $\sim0\fm18$ than the general trend defined by the M-type stars, whereas the S stars are all fainter than this trend, on average by $\sim0\fm18$. The number of C and S stars is too small to make a reliable linear least-squares fit to the data. Therefore, to fit their trend, we  calculated their mean offset from the overall fit and adopted the slope derived from the M stars. For this, we excluded the C star S~Cep, which is a very bright outlier, and the S star RW~And, which is a faint outlier. The coefficients of the Wesenheit index relations are presented in Table~\ref{tab:Wesenheit}. The carbon stars are, on average, not much brighter than the S stars in the DIRBE [2.2] band, as demonstrated by the $[2.2]$ versus $[1.25]-[2.2]$ colour-magnitude diagram displayed in Fig.~\ref{fig:DIRBE_CMD}. However, the C stars are considerably redder than the S stars in $[1.25]-[2.2]$ due to the enhanced circumstellar extinction of the carbon dust and by the appearance of the CN red band in the $J$ band in carbon rich atmospheres. This red colour leads to a relatively large correction in the Wesenheit $W_{[2.2],[1.25]-[2.2]}$ index, so that C stars are brighter than the S stars by $\sim0\fm35$ at a given period in the $W_{[2.2],[1.25]-[2.2]}$ versus $\log(P)$ diagram. Comparing our Galactic sample to the results for the LMC by \citet[][their Table~1]{2007AcA....57..201S}, we again find that the Galactic stars are fainter in the Wesenheit index by $\sim0\fm22$ than the LMC stars, both O- and C-rich.

\begin{table}
\caption{Relations of the DIRBE Wesenheit index vs $\log P$, following the form $W_{[2.2],[1.25]-[2.2]} = b\times(\log P-2.38) + a$.}
\label{tab:Wesenheit}
\centering
\begin{tabular}{lcc}
\hline\hline
Sample       & $b$                & $a$              \\
\hline
DIRBE gold   & $-5.003\pm0.027$   & $-8.054\pm0.004$ \\
S stars only & $-5.003\pm0.027^c$ & $-7.879\pm0.004$ \\
C stars only & $-5.003\pm0.027^c$ & $-8.229\pm0.004$ \\
\hline
\end{tabular}
\tablefoot{$^c$: Adopting the same slope as for the overall DIRBE gold sample.}
\end{table}

\subsection{Synthetic SEDs, period--temperature, period--bolometric-luminosity, and period--radius relations of Miras}\label{sec:synthSEDs}

\begin{figure}
\centering
\includegraphics[width=\linewidth]{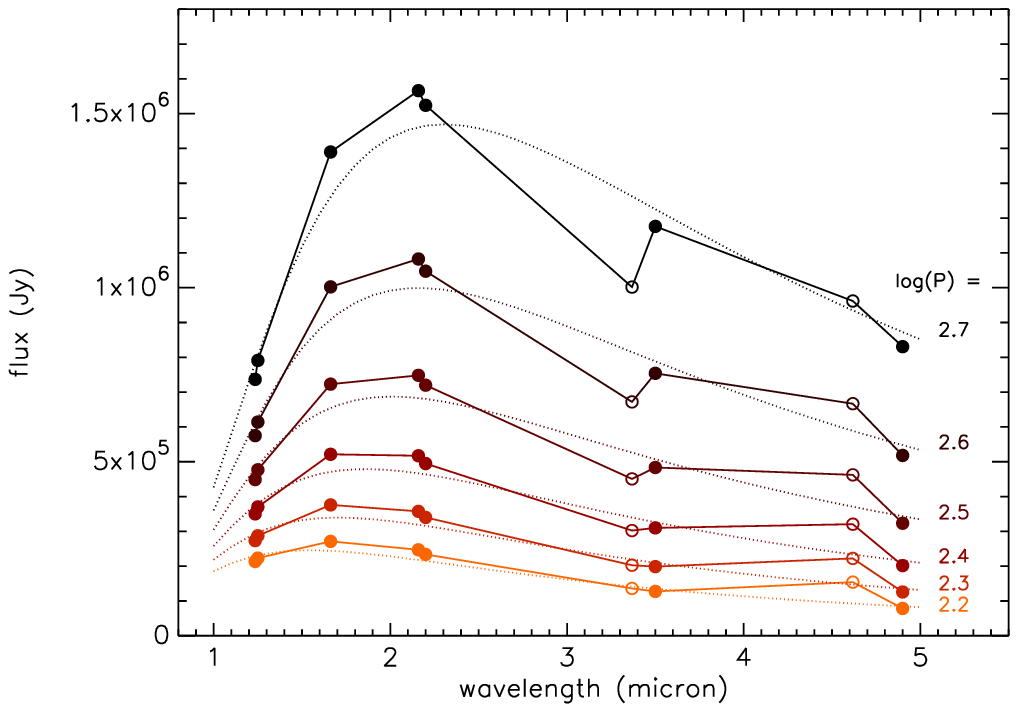}
\caption{Synthetic SEDs constructed with the PL relations given in Table~\ref{tab:PL relations}. The symbols connected by solid lines represent the fluxes derived for $\log(P) = 2.2$, 2.3, 2.4, 2.5, 2.6, and 2.7. The WISE bands are plotted as open symbols, while all other bands are plotted as filled symbols. The dotted lines are blackbody fits to each of the six synthetic SEDs, excluding the WISE bands.}
\label{fig:synthSEDs}
\end{figure}

With the help of the PL relations in the nine IR bands we established above, we were able to construct mean synthetic spectral energy distributions (SEDs) at any $\log(P)$ value of our Mira sample. These synthetic SEDs represent the average flux distribution of the sample Miras at this given period. We evaluated the PL relations in Table~\ref{tab:PL relations} at $\log(P) = 2.2$, 2.3, 2.4, 2.5, 2.6, and 2.7, which sample and represent the period distribution of our Mira stars well. The synthetic SEDs thus generated are shown in Fig.~\ref{fig:synthSEDs}. The symbols connected by solid lines represent the synthetic SEDs at the six $\log(P)$ values.

The nine photometric bands span the wavelength range around the maximum emission of the sample stars. This allowed us to make blackbody fits to the synthetic SEDs to derive additional (average) properties of Galactic Miras. The blackbody fits to the six synthetic SEDs are shown as dotted lines in Fig.~\ref{fig:synthSEDs}. The WISE bands, which we plot as open symbols in Fig.~\ref{fig:synthSEDs}, markedly deviate from the Planck curves: the W1 band at $\sim3.4\,\mu{\rm m}$ tends to fall below the blackbody fit, in particular at the long-period end, whereas the W2 band at $\sim4.6\,\mu{\rm m}$ tends to fall above the blackbody fit. Therefore, we decided to exclude both bands from the blackbody fits, even though including them would not significantly alter the results. As we are not aware of any (molecular) features in Mira spectra that could affect the flux in the two WISE bands so dramatically, we think that they deviate from the trend of the other bands because of remaining issues with reconstructing the flux of these bright, partially saturated objects.

\begin{figure}
\centering
\includegraphics[width=\linewidth]{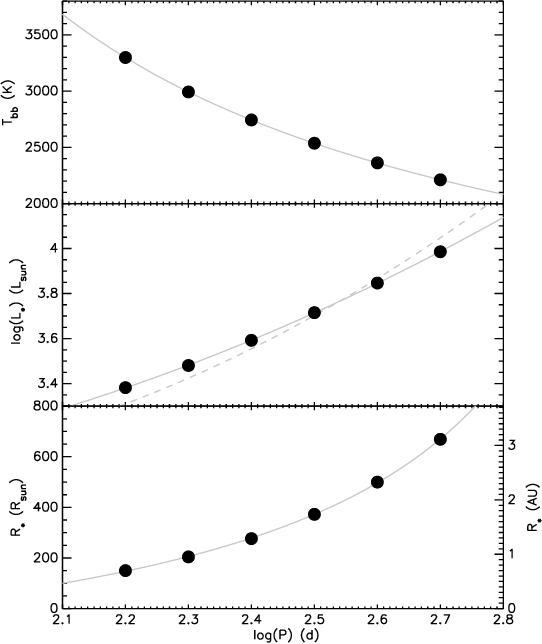}
\caption{Relations derived from the blackbody fits to the synthetic SEDs in Fig.~\ref{fig:synthSEDs}. {\it Top panel:} Blackbody temperature, $T_{\rm bb}$. {\it Middle panel:} Logarithm of the stellar luminosity in solar units. {\it Bottom panel:} Stellar radius in solar units and AU as a function of $\log(P)$. The solid grey lines show the relations given by Eq.~\ref{Eq:Tbb}, \ref{Eq:Lstar}, and \ref{Eq:Rstar}, and the dashed grey line in the middle panel is the PL relation derived from data reported by \citet{2000MNRAS.319..728W}.}
\label{fig:logP-Tbb-logL-R}
\end{figure}

The blackbody temperature, $T_{\rm bb}$, is a direct result of the fit of Planck curves to the synthetic SEDs. It varies between 3299\,K at $\log(P)=2.2$ and 2212\,K at $\log(P)=2.7$. We plot the best-fitting blackbody temperatures as a function of $\log(P)$ in the upper panel of Fig.~\ref{fig:logP-Tbb-logL-R}. The run of $T_{\rm bb}$ as a function of $\log(P)$ may be fit by a function of the form
\begin{equation}\label{Eq:oneoverx}
T_{\rm bb}=\frac{c}{\log(P) + a} + b.
\end{equation}
From a fit to $T_{\rm bb}$ at the six representative $\log(P)$ values, we find
\begin{equation}\label{Eq:Tbb}
    T_{\rm bb}=\frac{2655.0}{\log(P) - 1.3167} + 293.0.
\end{equation}
This fitting function is plotted as a grey line in the top panel of Fig.~\ref{fig:logP-Tbb-logL-R}. The individual points deviate by less than 0.5\,K from this fit, with a root mean square (rms) of 0.3\,K.

Furthermore, by integrating the Planck curve, we calculated the bolometric luminosity of the synthetic SEDs and thus estimated the average luminosity of Miras at a given pulsation period to establish a period--bolometric-luminosity relation. Deriving the bolometric luminosity of the stars from a blackbody fit necessarily neglects absorption by atoms, molecules, and circumstellar dust grains, as well as dust emission in the mid-IR. However, as our sample exclusively contains optically bright Miras, at least the dust absorption and emission should be relatively modest. We regard the resulting luminosity as a fair estimate of the average bolometric luminosity of optical Miras at a given period. The period--bolometric-luminosity relation following from our blackbody fits is shown in the middle panel of Fig.~\ref{fig:logP-Tbb-logL-R}. The bolometric luminosity varies between 2410\,$L_{\sun}$ at $\log(P)=2.2$ and 9671\,$L_{\sun}$ at $\log(P)=2.7$. The relation is almost linear, but adding a quadratic term significantly improves the fit. We fit a parabola of the form
\begin{equation}\label{Eq:Lstar}
    \log(L_*/L_{\sun})=3.708-1.256\log(P)+0.5035\log(P)^2
\end{equation}
to the data points at the six representative $\log(P)$ values. The data deviates by at most 22\,$L_{\sun}$ from the fit, with an RMS of 16\,$L_{\sun}$.

\citet{2000MNRAS.319..728W} determined bolometric magnitudes of a set of Miras from blackbody fits to near-IR observations in a very similar way as we did. We converted these to absolute bolometric magnitudes using the stars' distances given by \citet{2021AJ....161..147B} and added periods from the Karlsson for a few stars with missing periods. The \citet{2000MNRAS.319..728W} sample also indicates a non-negligible quadratic term in $\log(P)$. Excluding R~Cen and T~Sgr because of their very uncertain distances (and thus unrealistically high luminosities), we fitted the relation $\log(L_*/L_{\sun})=3.708-1.256\log(P)+0.5035\log(P)^2$ to the data. This relation is plotted as a dashed grey line in the middle panel of Fig.~\ref{fig:logP-Tbb-logL-R}. The agreement with our results is very good.

Finally, by applying the Stefan-Boltzmann law, $L=4 \pi R^2 T^4$, we also calculated the average stellar radii. The results of this are shown in the lower panel of Fig.~\ref{fig:logP-Tbb-logL-R}. For the six representative $\log(P)$ values, we find radii between $150\,R_{\sun}$ at $\log(P)=2.2$ and $669\,R_{\sun}$ (corresponding to 0.7 and 3.1\,AU, respectively) at $\log(P)=2.7$. To fit the data, we chose a function of the same form as Eq.~\ref{Eq:oneoverx}, which yielded the equation
\begin{equation}\label{Eq:Rstar}
    R_*/R_\sun=\frac{-534.5}{\log(P) - 3.208} - 382.8.
\end{equation}
The data deviate by at most 3\,$R_\sun$, with an RMS of 2\,$R_{\sun}$. Note that this is the stellar radius fulfilling the Stefan-Boltzmann law, given the temperature and luminosity from the blackbody fits. As the radius of these highly extended objects is hard to define, different definitions of the stellar radius exist, such as that of the (IR) continuum in interferometric observations \citep{2026A&A...711A.259J}. The above relations may be useful for deriving average stellar properties in the study of individual objects or populations of Miras.

\section{Testing the \Gaia parallax uncertainties}\label{sec:tests}

In this section, we present three different approaches to test whether the parallax uncertainties of AGB stars provided by the \Gaia DR3 catalogue are realistically estimated or if the uncertainties for this highly extended stellar type are underestimated. The first approach in Sect.~\ref{sec:PL-test} uses the RMS scatter around the PL relation in the DIRBE [2.2] band to test the uncertainties. The second approach in Sect.~\ref{sec:VLBI-test} takes a fresh look at the comparison with the parallaxes of AGB stars measured with VLBI methods. In doing so, we revisit the analysis by \citet{2022A&A...667A..74A}. The last approach, presented in Sect.~\ref{sec:47Tuc}, uses the LPVs in the globular cluster 47~Tuc to test if their parallax uncertainties are potentially underestimated.

\subsection{Estimating the parallax uncertainties from the PL($K$) relation}\label{sec:PL-test}

The following test uses the astrophysical condition that Mira variables follow a (nearly) linear relation between the logarithm of the pulsation period and their absolute brightness. The magnitude of the scatter around the Mira PL sequences then depends on the uncertainty of the distance measurement used to derive the absolute brightness and any uncertainty in the variable's mean brightness. Since we minimised the uncertainty of the mean brightness by using multi-epoch information, the scatter in the diagrams can be used to make an estimate of whether the parallax uncertainty is underestimated. For each star, we computed the quantity $f=\Delta M/\sigma_{m,\varpi}$, where $\Delta M$ is the difference between each star's $M_0$ magnitude and the linear fit (Table~\ref{tab:PL relations}) and $\sigma_{m,\varpi}$ is the uncertainty in absolute magnitude originating from the parallax distance uncertainty. If the parallax uncertainty given in the \Gaia catalogue is of the correct order, the standard deviation of $f$ should be close to 1. This quantity thus allows one to detect and quantify an underestimation of the parallax error. For the DIRBE gold sample in the [2.2] band, the standard deviation of $f$ is 2.77. This is the maximum factor by how much the parallax distance uncertainty could be underestimated. Other bands have similar factors $f$; in the DIRBE [3.5] band, it is 2.76, and in the $W1$ band, it is 2.74.

This simple estimate only yields an upper limit to the error inflation factor (EIF) because it does not take into account the intrinsic width of the PL sequence. As AGB stars experience thermal pulse (TP) cycles, which are violent and quasi-periodic ignitions of their He-burning shells, evolutionary models predict that the stars will cross the pulsation sequence (here, we investigate only the FM sequence) on slanted paths. Depending on the initial mass, each of these TP cycles takes thousands or tens of thousands of years, and we do not know in which phase of its TP cycle an individual star is. Therefore, we have to model the intrinsic width of the FM sequence to further constrain the EIF.

For this simulation, we used a small grid of AGB evolutionary model tracks calculated with the \texttt{COLIBRI} code of \citet{2013MNRAS.434..488M} with a near-solar metallicity of $[{\rm Fe}/{\rm H}]=-0.03$ and masses of $M=1.0$, 1.3, 1.5, 1.8, 2.0, 2.4, 2.6, 3.0, 4.0, and 5.0\,$M_{\sun}$. Based on these tracks, non-linear pulsation modes and periods were calculated with the analytic relations of \citet{2021MNRAS.500.1575T}. Only those parts of the tracks during which the model star pulsates in the FM were selected for the analysis. Each model time step $i$ was assigned a weight $w_{\rm i} \propto \Delta t_{\rm i}\times M^{-2.3}$, where $\Delta t_{i}$ is the time the model remains in that step, and the exponent $\alpha=-2.3$ corresponds to a 'Kroupa initial mass function' \citep{2001MNRAS.322..231K}. Furthermore, with this weighting, we assumed a constant star-formation rate in the solar neighbourhood. All weights were normalised to a maximum of 1.0. The model tracks come with absolute magnitudes in several photometric bands, including the 2MASS bands (but not the WISE bands, for example). Thus, we constructed a theoretical PL($K$) diagram with the non-linear pulsation periods and the $M_{K,\rm S}$ magnitude, which we treated as being identical to the DIRBE [2.2] band.

\begin{figure}
\centering
\includegraphics[width=\linewidth]{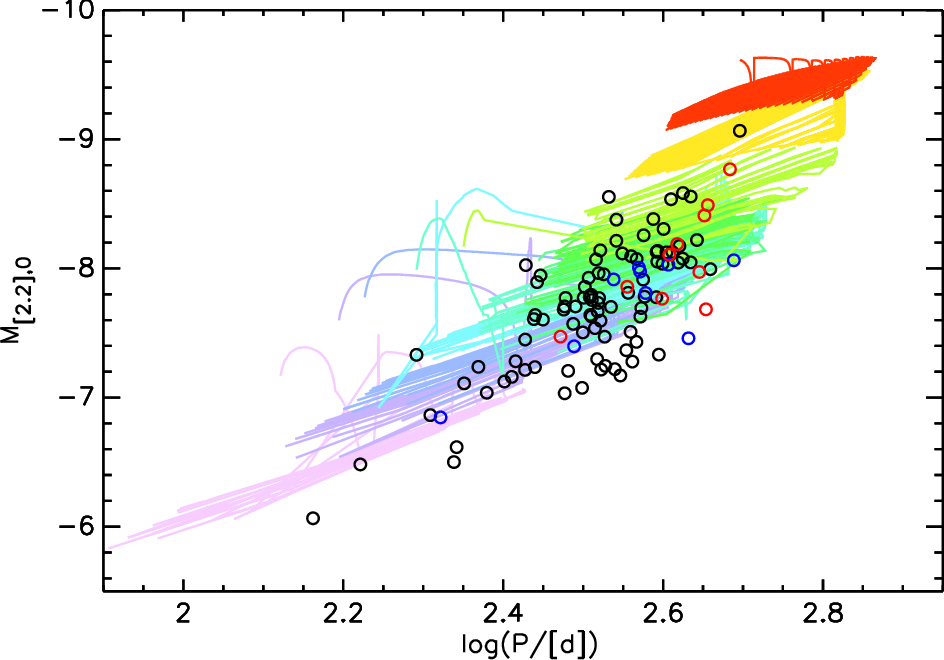}
\caption{Comparison of the FM \texttt{COLIBRI} model tracks with the DIRBE [2.2] gold sample in the PL($K$) diagram. The tracks are plotted as lines with colours according to their initial masses between 1.0\,$M_{\sun}$ (pink line, lower left) and 5.0\,$M_{\sun}$ (red, upper right). The DIRBE [2.2] gold sample stars are plotted as open circles with their colours distinguishing chemical spectral types (see Fig.~\ref{fig:DIRBE_PL2.2}).}
\label{fig:PK_model}
\end{figure}

In Fig.~\ref{fig:PK_model}, we compare the {FM} \texttt{COLIBRI} model tracks with the DIRBE [2.2] gold sample in the PL($K$) plane. The model grid fills the plane without significant gaps. We see that the extension of the tracks at a given period reasonably resembles the observed width of the PL($K$)-relation, and most stars in this diagram fall onto one of the model tracks. Note that the figure shows all time steps of the grid, irrespective of how much time the model star spends in each point; the few excursions to shorter periods are particularly short-lived. Any comparison with the magnitude scatter of a sample, such as in the LMC, must take into account the time spent at a given point.

In the next step, we constructed mean PL($K$) relations from the model tracks. For this bootstrapping method, we selected the same number of time steps (model stars) as we have observed stars in the DIRBE gold sample (110). A randomly selected model point $i$ is kept in the model ensemble if its weight $w_i$ is above a threshold formed by a uniformly drawn number $x$, $x\in[0,1]$. To realistically simulate the PL($K$) sequence, we added a normally distributed random scatter to $K_{\rm S,i}$, for which we adopted a $\sigma_{K,{\rm S}}$ with the same magnitude as a random one of the 110 observed parallax distance uncertainties of the DIRBE gold sample. This $\sigma_{K,{\rm S}}$ was artificially inflated with EIFs between 1.0 and 5.0, where the latter was suggested by \citet{2022A&A...667A..74A}. The photometric uncertainty, also derived from the observations of the gold sample, was added to this distance-based uncertainty in quadrature. However, the photometric uncertainty is clearly inferior to the parallax uncertainty. Thirty bootstrapping ensembles of 110 model steps were created for each chosen EIF, and the statistical scatter around a linear $\log P - M_{K,S}$ fit of each ensemble was evaluated. Note that this simulation does not take into account a distribution in the metallicity of the stars, the uncertainty in interstellar extinction, or other factors. Thus, this test yields an upper limit of the EIF.

\begin{figure}
\centering
\includegraphics[width=\linewidth]{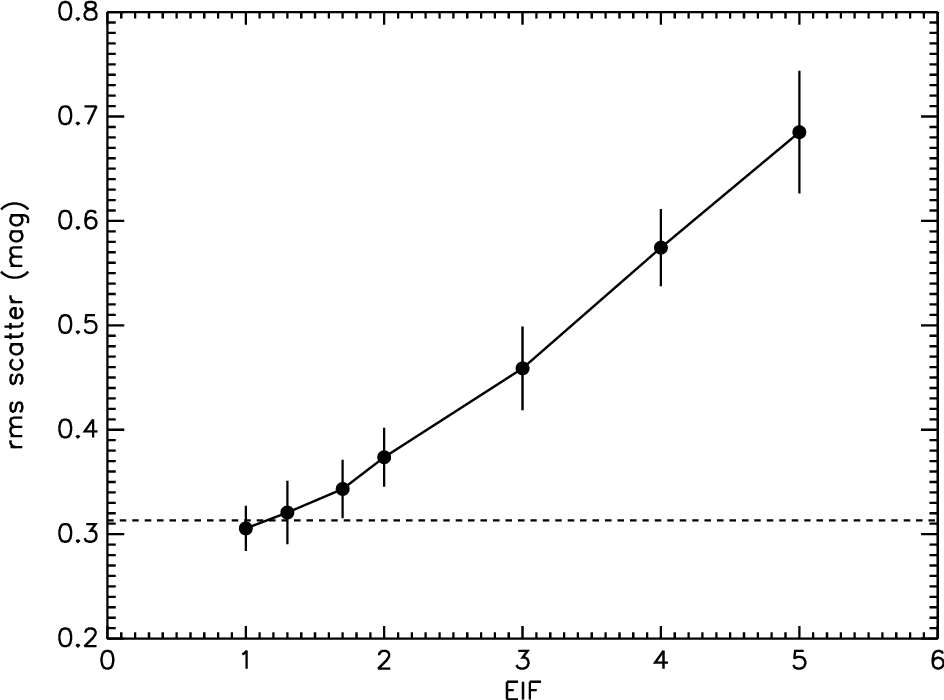}
\caption{Root mean square scatter around the simulated PL($K$) relation of the \texttt{COLIBRI} evolutionary grid as a function of the EIF. The error bars indicate the $1\sigma$ range of the RMS scatter of the 30 realisations for each EIF. The horizontal dashed line indicates the observed scatter of 0\fm313 around the PL relation of the DIRBE [2.2] band shown in Fig.~\ref{fig:DIRBE_PL2.2}.}
\label{fig:EIF-analysis}
\end{figure}

The result of our exercise is shown in Fig.~\ref{fig:EIF-analysis}, which plots the RMS scatter around the simulated PL($K$) relation of the \texttt{COLIBRI} evolutionary grid as a function of the EIF. It is not possible to precisely pinpoint how large the EIF of the AGB \Gaia parallax uncertainties is. Nevertheless, we see from Fig.~\ref{fig:EIF-analysis} that the \Gaia DR3 parallax uncertainties of the DIRBE gold sample (dashed line) are compatible with the scatter around the simulated PL($K$) relation of the \texttt{COLIBRI} evolutionary grid, which means that the observed scatter is roughly consistent with the uncertainties given in the \Gaia catalogue. The uncertainties are also consistent with an EIF of 1.3 and marginally consistent with an EIF of 1.7. However, significantly larger EIFs can be safely excluded. At an EIF of 5, the PL relation can hardly be recognised as a sequence of increasing brightness with an increasing pulsation period.

We caution, however, that our method to estimate the EIF from the scatter around the Mira PL($K$) relation has shortcomings. The non-linear pulsation models of \citet{2021MNRAS.500.1575T} have been tested on observations of LPVs in the MCs, but not for the higher metallicity of the Milky Way galaxy. Numerical tests by \citet{2021MNRAS.500.1575T} have shown that the non-linear pulsation periods derived from the analytic expressions predict those from the actual time series within a 10\% error in more than 80\% of the cases. This means that in some cases they deviate by more than 10\% in the pulsation period. The dominant pulsation mode is selected based on the growth rates of the modes, and there may be considerable uncertainty in these. Furthermore, the models might not be so well constrained at high masses ($M\gtrsim 4M_{\sun}$) as described in \citet{2021MNRAS.500.1575T}. From Fig.~\ref{fig:PK_model} we may, however, safely assume that the typical mass of solar neighbourhood Miras is between 1 and 3 $M_{\sun}$. Finally, the models of \citet{2021MNRAS.500.1575T} approximate turbulent mixing in the envelope of AGB stars with the classical mixing length parameter. This mixing length parameter is neglected altogether in the analytic expressions used to calculate the non-linear periods of the model grid employed here. All these shortcomings may affect both the absolute values of brightnesses and  periods and the range of variation in these two parameters along the PL relation. Since our main interest is on the intrinsic scatter around this relation rather than its absolute location in the PL-plane, only the latter is relevant here. However, models do not  yet allow the effect of these shortcomings on our conclusions to be quantified.

Independently of the models, we can estimate from the magnitude of the RMS scatter of stars around the PL relation whether bright stars have more substantially underestimated parallax uncertainties than stars of lower brightness. We made this estimate with the unTimely $W1$ gold sample due to its larger size. We divided the gold sample at the median \Gaia $G$ band magnitude (8\fm768) into two equal halves of 156 stars. The bright gold sample has an average relative distance uncertainty of 0.050, and the faint gold sample has 0.072. The RMS scatter around the PL($W1$) relation of the two groups is 0\fm313 and 0\fm339, respectively. Thus, the brighter stars have a smaller RMS scatter, but they also have more precise parallaxes on average. The numbers do not suggest that the parallax uncertainties of the bright stars could be more severely underestimated than those of the faint stars.

\subsection{Comparison of PL relations based on \Gaia and VLBI parallaxes}\label{sec:VLBI-test}

A direct comparison of parallax distances determined from \Gaia and Very Long Baseline Interferometry (VLBI) of maser-emitting AGB stars can provide insight as to whether the combined uncertainty estimates are realistic. The parallax difference $\varpi_{\rm Gaia}-\varpi_{\rm VLBI}$, normalised by their quadratically combined uncertainties $\sqrt{\sigma_{\rm Gaia}^2 + \sigma_{\rm VLBI}^2}$, has been inspected for samples of AGB stars in the past to estimate parallax zero-point offsets (ZPOs) and EIFs \citep{2018evn..confE..43V,2022A&A...667A..74A}. \citet{2022A&A...667A..74A} provided arguments as to why the VLBI parallaxes and their associated uncertainties should be robust and highly accurate. They found an EIF of 4.20 for their entire AGB sample and an EIF of up to 5.44 for the optically brightest stars. However, from a purely theoretical point of view, we cannot know if the assumption of robustness of the VLBI parallax uncertainties is warranted. For example, it is unknown how well the assumption of structural invariability of the maser emission is fulfilled or how much a binary companion impacts the maser emitting circumstellar environment \citep{2026Galax..14...26N}. The EIF may actually have to be distributed to both the \Gaia and the VLBI parallax uncertainties, or systematic uncertainties may affect the analysis.

We can gain additional insight into the robustness of \Gaia and VLBI parallaxes and their uncertainties by inspecting PL diagrams of pulsating AGB stars constructed with the respective parallax distances. Such a diagram has been constructed by \citet[][their Fig.~9]{2022A&A...667A..74A} using VLBI parallaxes. However, we identified two problems with that diagram. First, the bolometric magnitudes calculated by \citet{2022A&A...667A..74A} mainly rely on single-epoch 2MASS photometry. Most optical AGB stars have their emission maximum around the $J$, $H$, and $K$ bands. Thus, these bands have a strong impact on the bolometric magnitudes, which will therefore exhibit increased scatter due to the non-negligible variability of the stars. To overcome this limitation, we used the mean unTimely $W1$ photometry to construct a PL($W1$) diagram; the typically 16 epochs of WISE observations significantly reduce the variability-induced scatter. Second, we found that the VSX pulsation periods \citep{2006SASS...25...47W} used by \citet{2022A&A...667A..74A} in some cases significantly disagree with the period in the databases we used. Therefore, we redetermined all periods of our sample stars based on light curves available from different sources. Where available, we adopted periods as listed by T.\ Karlsson. If that was not available, we determined the period from photometry available in the Kamogata/Kiso/Kyoto Wide-field Survey \citep[KWS;][]{Maehara2014}, the AAVSO, or The All Sky Automated Survey \citep[ASAS;][]{Pojmanski2002} by a Fourier analysis of the light curves with the Period04 program \citep{2004IAUS..224..786L}. NSV~17351 has such a long pulsation period that the cadence of WISE observations is high enough to determine its period directly from the $W1$ light curve with a simple sine fit. Finally, the period of the Mira OZ~Gem is adopted from \citet{2018AJ....156..241H}, who provide a light curve from which the reliability of the period was checked. We list the newly determined pulsation periods of the sample of AGB stars with VLBI parallaxes in Table~\ref{tab:Per-corr} and compare them to the VSX period.

We added the three Miras R~Cas, T~UMa, and U~CVn to the sample of \citet{2022A&A...667A..74A}. Their VLBI parallaxes have been adopted from \citet{2003A&A...407..213V}, \citet{2018IAUS..336..365N}, and the \citet{2020PASJ...72...50V}, respectively. Additionally, for NSV~17351, we adopted the new VLBI parallax of $\varpi_{\rm VLBI}=0.247\pm0.035$\,mas from \citet{2023PASJ...75..529N}.

\begin{table}
\caption{Improved pulsation periods of AGB stars with measured VLBI parallaxes, plotted in Fig.~\ref{fig:Gaia-VLBI_PL-W1}.}
\label{tab:Per-corr}
\centering
\begin{tabular}{lrrrl}
\hline\hline
Object    &	Var.    &	$P$ (VSX) & $P$ (new)  & Source \\
          &         &     (d)     &    (d)     &     \\
\hline
AP Lyn	  & 	M   & 	730.0     &  433.9     & KWS \\
BX Cam	  &     M   & 	486.0     &  439.4     & AAVSO \\
BX Eri	  &     SR  &  	165.0     &  165.9     & KWS \\
FV Boo	  &     M   &   313.0     &  306.73    & KWS \\
HS UMa	  &     LB  & 	  0.0     &  331.1:    & KWS \\
HU Pup	  &     SRa & 	238.0     &  828.0     & ASAS \\
NSV 17351 & 	M   & 	680.0     &  1108.0    & unTimely \\
OZ Gem	  &     M   & 	598.0     &  603.72    & H+2018 \\
QX Pup	  &     M   & 	551.0     &  546.8     & AAVSO \\
R Aqr	  &     M   & 	387.0     &  382.14    & Karlsson \\
R Cas	  &     M   &   430.6     &  431.00    & Karlsson \\
R Cnc	  &     M   & 	357.0     &  364.86    & Karlsson \\
R Hya	  &     M   & 	380.0     &  355.43    & Karlsson \\
R Peg	  &     M   & 	378.1     &  377.93    & Karlsson \\
R UMa	  &     M   & 	301.62    &  301.35    & Karlsson \\
RR Aql	  &     M   & 	396.1     &  397.15    & Karlsson \\
RT Vir	  &     SRb & 	157.9     &  158.17    & KWS \\
RW Lep	  &     SRa & 	149.9     &  289.38    & KWS \\
RX Boo	  &     SRb & 	158.0     &  163.45    & KWS \\
S CrB	  &     M   & 	360.26    &  359.79    & Karlsson \\
S Crt	  &     SRa & 	155.0     &  305.59    & KWS \\
S Ser	  &     M   & 	371.84    &  367.08    & Karlsson \\
SV Peg	  &     SRb & 	144.6     &  318.34    & KWS \\
SY Aql	  &     M   & 	355.92    &  357.36    & Karlsson \\
SY Scl	  &     M   & 	411.0     &  415.83    & KWS \\
T Lep	  &     M   & 	372.0     &  370.57    & Karlsson \\
T UMa	  &     M   &   256.35    &  256.79    & Karlsson \\
U CVn	  &     M   &   341.6     &  341.00    & Karlsson \\
U Her	  &     M   & 	404.0     &  401.29    & Karlsson \\
U Lyn	  &     M   & 	433.6     &  434.93    & Karlsson \\
UX Cyg	  &     M   & 	569.0     &  577.86    & Karlsson \\
V637 Per  &     SR  & 	  0.0     &  182.6:    & KWS \\
V837 Her  &     M   & 	514.0     &  525.93    & KWS \\
W Leo	  &     M   & 	391.75    &  386.43    & Karlsson \\
X Hya	  &     M   & 	299.5     &  298.20    & Karlsson \\
Y Lib	  &     M   & 	277.0     &  275.14    & Karlsson \\
\hline
\end{tabular}
\tablefoot{Column~1: Object name. Column~2: Variability type: M: Mira, SR/SRa/SRb: semi-regular variables, LB: Irregular variables. Column~3: Pulsation period as reported in VSX \citep{2006SASS...25...47W}. Column~4: New period determination (this work). Column~5: Source for the data used for the new period determination: KSW: Kamogata/Kiso/Kyoto Wide-field Survey \citep{Maehara2014}, AAVSO: \url{https://www.aavso.org/}, ASAS: The All Sky Automated Survey \citep{Pojmanski2002}, unTimely: \citet{2023AJ....165...36M}, H+2018: \citet{2018AJ....156..241H}, Karlsson: \citet{2013JAVSO..41..348K}.}
\end{table}

\begin{figure}
\centering
\includegraphics[width=\linewidth]{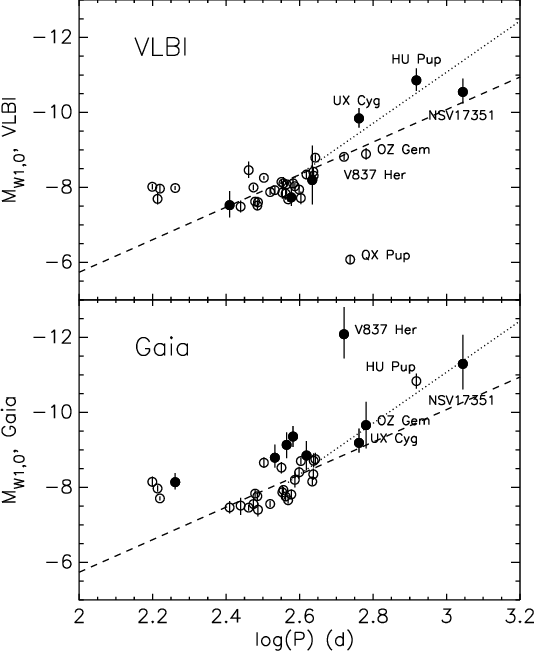}
\caption{Period--luminosity ($W1$) diagram of the sample stars from \citet{2022A&A...667A..74A}. {\it Top panel:} $M_{{\rm W1},0}$ based on the VLBI parallax distance. {\it Bottom panel:} $M_{{\rm W1},0}$ based on the \Gaia distance. Objects with $\log(P)\ge2.65$ are identified by their names. Open symbols represent stars with $\sigma_{\varpi}/\varpi<0.1$; filled symbols are for those with $\sigma_{\varpi}/\varpi\ge0.1$. The dashed line is the relation determined from the gold sample in Fig.~\ref{fig:unTimely_PLW1} (see Table~\ref{tab:PL relations}), and the dotted line is the steeper relation for stars with $\log(P)\ge2.57$.}
\label{fig:Gaia-VLBI_PL-W1}
\end{figure}

The PL diagrams based on the VLBI and \Gaia parallaxes are presented in Fig.~\ref{fig:Gaia-VLBI_PL-W1}, where the upper panel displays the results for the VLBI and the lower panel those for the \Gaia. The comparison to the PL($W1$) relation determined from the gold sample displayed in Fig.~\ref{fig:unTimely_PLW1} (dashed line in Fig.~\ref{fig:Gaia-VLBI_PL-W1}) confirms that the stars form a very well defined Mira sequence, especially when using the VLBI parallaxes, but it is also well behaved when adopting the \Gaia parallaxes. By comparing the VSX periods with the newly determined periods in a PL diagram we concluded that the new periods also significantly reduce the scatter around a VLBI-based PL relation. In this way, obvious outliers and overtone-mode pulsators can be clearly identified (see below). There is no need to exclude many of the stars excluded by \citet{2022A&A...667A..74A} from determining the PL relation when using the improved pulsation periods.

It is worth inspecting the few outliers in the relation. The four stars with $\log(P)$ near 2.2, namely BX~Eri, RT~Vir, RX~Boo, and V637~Per, are semiregular variables (Table~\ref{tab:Per-corr}) and thus likely located on the first overtone PL relation. All other nominal non-Miras in the sample (HS~UMa, HU~Pup, RW~Lep, S~Crt, and SV~Peg) appear to be FM pulsators. QX~Pup does not have a counterpart in the \Gaia catalogue and therefore does not have a \Gaia parallax. With its VLBI parallax, it is much below the PL sequence formed by the FM pulsators. QX~Pup is also known as the Rotten Egg Nebula and is thought to be a protoplanetary nebula. Probably, it is so much dust-obscured that even the $W1$ band is significantly extinct. Because of the lack of a \Gaia parallax, it is not included in the comparison of the PL relations and the estimation of the EIF (see below), and also \citet{2022A&A...667A..74A} excluded it from determining their PL relation. UX~Cyg is found somewhat above the PL relation when using its VLBI parallax, but it is close to the steeper relation found for the stars with $\log(P)\ge2.57$ in Fig.~\ref{fig:unTimely_PLW1} (dotted line in Fig.~\ref{fig:Gaia-VLBI_PL-W1}). The AAVSO V band amplitude of $\sim6\fm0$ safely places UX~Cyg in the Mira class of variables. Also, HU~Pup is significantly above the PL relation, both when using the \Gaia and VLBI parallaxes, but fits well with the steeper slope. It has a very long period of 828\,d as we checked from the ASAS light curve, but it is not a Mira. Thus, possibly, it is a supergiant. Such massive stars are shifted to shorter periods and have amplitudes smaller than typical Miras \citep{2019A&A...631A..24L}. It is thought that these stars also pulsate in the FM.

The longest-period star, NSV~17351, deserves special attention. It is an OH/IR star with a period of 1108 days determined from the unTimely $W1$ light curve. \citet{2023PASJ...75..529N} report a period of $1122\pm24$\,d based on H$_2$O maser flux measurements. The \Gaia parallax ($\varpi_{\rm Gaia}=0.0884\pm0.1468$\,mas) is ill-defined, but the distance of $\sim5.7$\,kpc derived by \citet{2021AJ....161..147B} with the help of Bayesian statistics seems to be acceptable. When adopting the VLBI parallax \citep{2023PASJ...75..529N}, NSV~17351 is consistent with being on the PL($W1$) relation defined by the gold sample. There is a long-standing discussion in the literature about whether OH/IR stars at long periods ($P\gtrsim1000$\,d) follow the PL relations of shorter-period Miras \citep[e.g.][]{1991MNRAS.248..276W,2022A&A...659A.145G,2024IAUS..376..328E}. The result for NSV~17351 argues in favour of the PL relation extending to these long-period OH/IR stars.

A PL diagram of stars in the VLBI-sample (Fig.~\ref{fig:Gaia-VLBI_PL-W1}) can give us clues about the reliability of the \Gaia and VLBI parallaxes and their uncertainties. It is particularly instructive to inspect the stars with the larger parallax uncertainties, $\sigma_{\varpi}/\varpi\ge0.1$, represented by filled symbols in Fig.~\ref{fig:Gaia-VLBI_PL-W1}. In the PL($W1$) diagram based on the VLBI parallaxes (upper panel), we see that those stars, even if they are few, do not systematically deviate from the PL relation of the gold Mira sample indicated by the dashed line (we remind that HU~Pup might be located above the PL($W1$) relation intrinsically and four stars are likely first-overtone pulsators). Looking at the PL($W1$) diagram based on the \Gaia parallaxes, on the other hand, we find that almost all stars deviating from the PL sequence significantly show $\sigma_{\varpi}/\varpi\ge0.1$, and all of them are shifted to higher luminosity, V837~Her even by a large margin. This indicates that their distances are systematically overestimated. Including these stars in a comparison with the VLBI parallaxes, as done by \citet{2022A&A...667A..74A}, will necessarily increase the estimated EIF.

On the other hand, we find indications that some of the VLBI parallax uncertainties could be underestimated. As noted above, here we adopted the parallax measurement of NSV~17351 from \citet{2023PASJ...75..529N}. The uncertainty reported by these authors is 3.5 times larger than the one used by \citet{2022A&A...667A..74A}. Six of the \citet{2022A&A...667A..74A} sample stars have VLBI parallaxes with relative uncertainties of 2\% or less. In contrast, only two of the 312 unTimely $W1$ gold sample stars have relative \Gaia parallax uncertainties slightly below 2\%. It is advisable to check if the small VLBI uncertainties are indeed realistic \citep{2020PASJ...72...50V}.

As previously done by \citet{2018evn..confE..43V} and \citet{2022A&A...667A..74A}, we useed this VLBI sample to test the reliability of the uncertainty estimates. \citet{2022A&A...667A..74A} assumed the VLBI parallax uncertainties to be robust and defined a one-sided EIF that applies to the \Gaia parallax uncertainties only such that the quantity
\begin{equation}\label{Eq:EIF-1side}
    \frac{\Delta\varpi}{\sigma_{\varpi,{\rm tot}}}=\frac{\varpi_{\rm Gaia} - \varpi_{\rm VLBI}}{\sqrt{\left({\rm EIF}\cdot\sigma^{\rm Gaia}_{\varpi}\right)^2+\left(\sigma^{\rm VLBI}_{\varpi}\right)^2}}
\end{equation}
has a distribution with a standard deviation of 1.0. Here, $\varpi_{\rm Gaia}$ and $\sigma^{\rm Gaia}_{\varpi}$ are the \Gaia parallax and its associated uncertainty, and $\varpi_{\rm VLBI}$ and $\sigma^{\rm VLBI}_{\varpi}$ are the same for the VLBI measurements. As argued at the beginning of this section, we do not know a priori that the VLBI parallaxes are estimated correctly, and some of them could be underestimated. We therefore also defined a symmetric or two-sided EIF $\xi$ that applies to both \Gaia and VLBI parallax uncertainties such that the quantity
\begin{equation}\label{Eq:EIF-2side}
    \frac{\Delta\varpi}{\sigma_{\varpi,{\rm tot}}}=\frac{\varpi_{\rm Gaia} - \varpi_{\rm VLBI}}{\sqrt{\left(\xi\cdot\sigma^{\rm Gaia}_{\varpi}\right)^2+\left(\xi\cdot\sigma^{\rm VLBI}_{\varpi}\right)^2}}
\end{equation}
has a distribution with a standard deviation of 1.0. We did not apply a ZPO to correct the \Gaia parallaxes ($\varpi^{\rm corr}_{\rm Gaia}=\varpi_{\rm Gaia}-{\rm ZPO}$) because we saw that this is strongly driven by the two nearest stars, R~Hya and RX~Boo ($d\approx150$\,pc), which both have considerably larger VLBI parallaxes than \Gaia parallaxes. This is most probably the reason why the ZPO determined by \citet[][$-0.131$\,mas]{2022A&A...667A..74A} for their AGB star sample has a significantly larger absolute value than those found by \citet[][$-0.021$\,mas]{2021A&A...649A...2L} or \citet[][$-0.039$\,mas]{2021A&A...654A..20G}, which are negligible.

We calculated the one-sided and the symmetric EIF for different subsamples to check how they vary with other parameters; the results are collected in Table~\ref{tab:EIFs}. Motivated by the results of the PL diagram (Fig.~\ref{fig:Gaia-VLBI_PL-W1}), we chose to restrict the sample to stars with a relative uncertainty $\sigma_{\varpi}/\varpi\le0.1$ in both the \Gaia and VLBI catalogues. This yielded a sample of 22 stars. The one-sided EIF (Eq.~\ref{Eq:EIF-1side}) of that sample is 4.59, and the two-sided EIF (Eq.~\ref{Eq:EIF-2side}) is 2.40. However, this value is strongly driven by the numbers for U~Her and RW~Lep, which have parallaxes that significantly deviate between \Gaia and VLBI observations. If we exclude them in the calculation of the EIFs, the factors decrease to 2.27 and 1.89, respectively.

\begin{table*}
\caption{One-sided (EIF), two-sided ($\xi$), and heteroscedastic Gaussian variance-components model ($\xi_{\rm G}$ and $\xi_{\rm V}$) error inflation factors to be applied to the parallax uncertainties to bring \Gaia and VLBI parallaxes into statistical agreement, for different selections of stars.}
\label{tab:EIFs}
\centering
\begin{tabular}{lccccc}
\hline\hline
Sample                                                         & N  & EIF  & $\xi$ & $\xi_{\rm Gaia}$ & $\xi_{\rm VLBI}$ \\
\hline
$\sigma_{\varpi}/\varpi\le0.1$                                 & 22 & 4.59 & 2.40 & $1.35\pm0.78$ & $3.44\pm0.95$ \\
$\sigma_{\varpi}/\varpi\le0.1\setminus\{{\rm U~Her, RW~Lep}\}$ & 20 & 2.27 & 1.89 & $2.13\pm0.58$ & $1.46\pm1.10$ \\
$G\le7\fm922\setminus\{{\rm U~Her, RW~Lep}\}$ ('bright')       & 16 & 2.69 & 2.17 & \dots         & \dots         \\
$G>7\fm922\setminus\{{\rm V837~Her}\}$ ('faint')               & 16 & 1.76 & 1.64 & \dots         & \dots         \\
\hline
\end{tabular}
\tablefoot{Column 1: Definition of the subsample. Column~2: Number of stars in that subsample. Column~3: One-sided EIF as defined by Eq.~\ref{Eq:EIF-1side}. Column~4: Two-sided EIF as defined by Eq.~\ref{Eq:EIF-2side}. Columns~5 and 6: Heteroscedastic Gaussian variance-components model inflation factors for \Gaia and VLBI, respectively.
}
\end{table*}

Similar to \citet{2022A&A...667A..74A}, we considered the bright and faint stars separately. We split the sample in half at the median $G$ band magnitude of $7\fm922$. Besides U~Her and RW~Lep (see above), we also excluded V837~Her from the faint sample because not only do the \Gaia and VLBI parallaxes differ significantly (0.1771\,mas and 1.09\,mas, respectively) but also because its VLBI parallax uncertainty is exceptionally small (0.01\,mas). As mentioned, QX~Pup has no \Gaia parallax. This yielded 16 stars in both the bright and faint subsamples. Both the one-sided and the two-sided EIFs are considerably smaller in the faint sample than in the bright sample. 

{\citet{2026Galax..14...26N} demonstrated that \Gaia parallax uncertainties vary in a considerably smaller range than those of VLBI. This fact can be exploited with a heteroscedastic Gaussian variance-components model, estimated by restricted maximum likelihood, to separately estimate the two EIFs, called $\xi_{\rm Gaia}$ and $\xi_{\rm VLBI}$ here. The optimisation yields well-constrained values only for the first two subsamples, listed in columns~5 and 6 in Table~\ref{tab:EIFs}. For the whole sample, the factors are $\xi_{\rm Gaia}=2.70\pm0.48$ and $\xi_{\rm VLBI}=2.58\pm0.67$. Experiments show that the results can change significantly on including or excluding individual stars. Nevertheless, this underscores that an inflation of only the \Gaia parallax uncertainties is not realistic.}

In summary, we confirm the results by \citet{2022A&A...667A..74A} that the fainter stars require smaller EIFs be applied to their parallax uncertainties to bring the parallaxes into statistical agreement. In particular, the one-sided EIF of the faint sample is only 1.76, which is considerably smaller than the EIFs reported by \citet[][their Table~2]{2022A&A...667A..74A}. However, we caution that the number of AGB stars that have parallaxes measured by \Gaia and VLBI is still very small, thus generally limiting the conclusions that can be drawn from such comparisons, and that a few individual stars strongly impact the results.

\subsection{\Gaia parallaxes of 47~Tuc AGB variables}\label{sec:47Tuc}

The reliability of \Gaia parallax uncertainties can also be inspected with a sample of AGB stars that are members of a stellar system and are thus at an essentially identical distance. The MCs host a large population of AGB stars; unfortunately, they are too distant to reliably measure parallaxes. Galactic open clusters are too sparsely populated to host sufficient numbers of stars in the sort-lived AGB phase \citep[see][for a recent collection of AGB stars in Galactic open clusters]{2022ApJS..258...43M}. Only nearby Galactic globular clusters are good candidates for assembling a large enough sample of AGB stars to test their \Gaia parallaxes. The globular cluster 47~Tucanae (47~Tuc) hosts a significant population of LPVs on the AGB that has been extensively studied.

In using stars from a Galactic globular cluster, we tested stars with a lower metallicity of $[{\rm M}/{\rm H}]=-0.78$ \citep{2009ApJ...694.1498M} and lower average mass than what is expected for the solar neighbourhood. Therefore, in this case, we studied objects with a smaller radius. Although the results from the cluster stars may not be directly applicable to the solar neighbourhood Miras, this test helps to derive a better understanding of the EIF for AGB stars.

We adopted the collection of 47~Tuc LPVs by \citet{2005A&A...441.1117L} to quantitatively test the reliability of their \Gaia parallax uncertainties. A qualitative test has already been performed by \citet[][their Fig.~33]{2023A&A...674A..15L}, indicating that the parallax uncertainties are at least realistic. From the 42 stars listed in Table~1 of \citet{2005A&A...441.1117L}, we discarded the star LW14, because no parallax is listed in the DR3, and LW12, because its DR3 parallax is about a factor of two larger than the average of all other stars, more than $5\sigma$ away from the cluster average, and thus it could be a foreground star. On the other hand, we retained V19 that was speculated by \citet{2005A&A...441.1117L} to be a foreground star; its parallax is fully compatible with being a cluster member. We also added V17, which is discussed in the main text of \citet{2005A&A...441.1117L} but is missing from their Table~1. Thus, we have a sample of 41 LPVs in 47~Tuc available for our parallax uncertainty test. Their \Gaia parallaxes are about a factor of ten larger than the associated uncertainties.

\begin{figure}
\centering
\includegraphics[width=\linewidth]{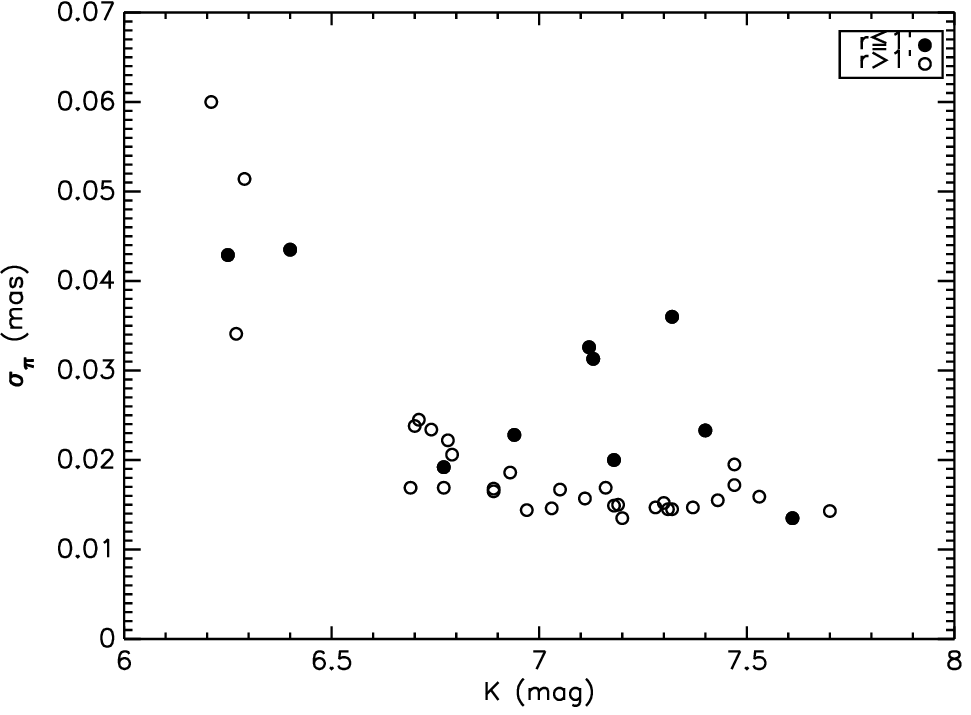}
\includegraphics[width=\linewidth]{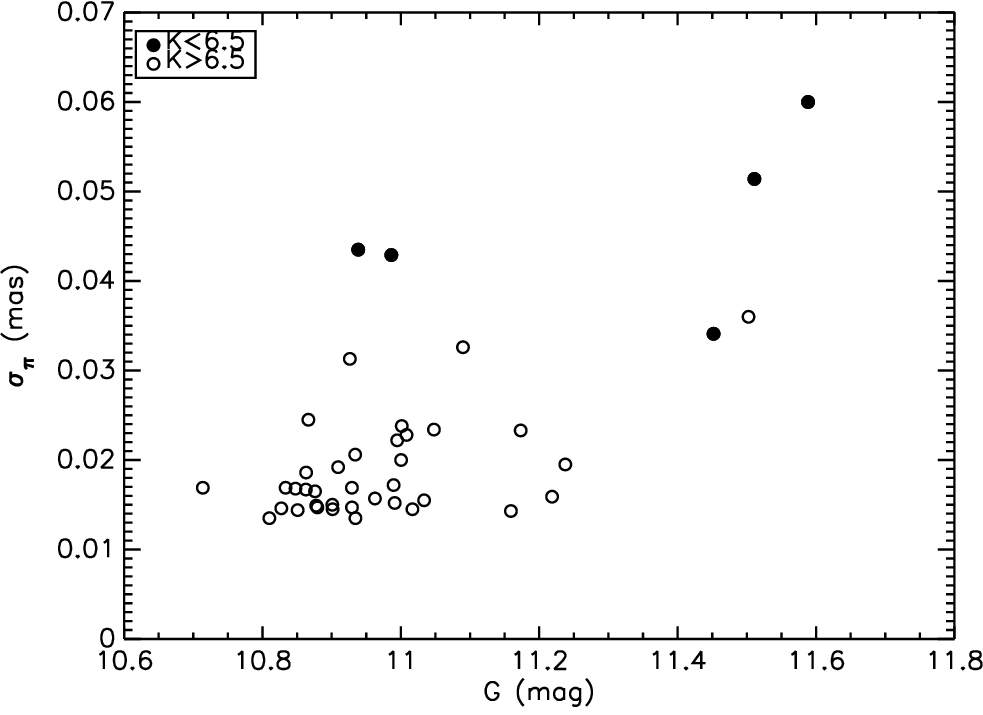}
\includegraphics[width=\linewidth]{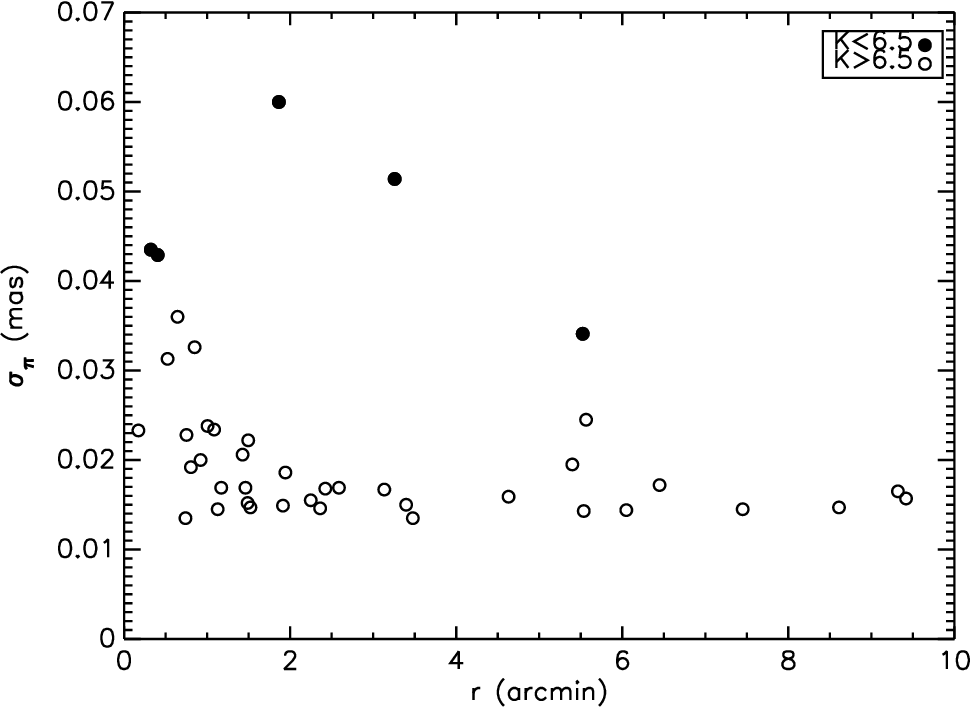}
\caption{\Gaia parallax uncertainties of 47~Tuc LPVs. {\it Upper panel:} Uncertainty as a function of $K$ band magnitude, which is proportional to the surface gravity $\log g$. Filled symbols show stars closer than $1\arcmin$ to the cluster centre; open symbols represent more distant stars. Brighter (i.e. more extended) stars have a larger parallax uncertainty because of stronger surface brightness variation. {{\it Middle panel:} Uncertainty as a function of \Gaia $G$ band magnitude.} {\it Lower panel:} Uncertainty as a function of angular distance from the cluster centre $r$. Filled symbols represent stars brighter than $K=6\fm5$, and open symbols represent stars fainter than this limit. The half-light radius of 47~Tuc is $\sim2.9\arcmin$. The uncertainty is enhanced close to the cluster centre because of the higher stellar density.}
\label{fig:47Tuc-test}
\end{figure}

As the size of convective cells on the surface increases with decreasing surface gravity, $\log g$ \citep{2018Natur.553..310P}, and since the parallax uncertainty for AGB stars is likely affected by large convective cells \citep{2018A&A...617L...1C}, we expect that the parallax uncertainty of 47~Tuc LPVs increases with decreasing surface gravity. The surface gravity decreases as a star becomes more luminous and extended. Thus, we expect the parallax uncertainty to increase with the $K$ brightness, which is a good proxy for the overall luminosity. The mass loss along the AGB will enhance the effect, but should play a minor role. In the upper panel of Fig.~\ref{fig:47Tuc-test}, \Gaia parallax uncertainties of 47~Tuc LPVs are plotted as a function of their mean $K$ band magnitude, adopted from \citet{2005A&A...441.1117L}. We can see that the brightest stars indeed have significantly increased parallax uncertainties. In addition, the fainter stars appear to have parallax uncertainties that increase slowly with $K$-band brightness.

However, it is not clear whether the large parallax uncertainties are primarily caused by the photocentre wobble of the AGB stars or not; \Gaia's deteriorating performance for very red sources or certain $G$ band magnitude ranges could play a significant role. \citet[][their Appendix C]{2024A&A...690A.125B} suggest that, from a comparison with normal M giants, the parallax uncertainty budget in Miras is indeed dominated by the photocentre shift due to the huge AGB convection cells, but the samples span different colour ranges. On the other hand, \citet{2023MNRAS.520.3510K} does not detect such shifts in red supergiant stars, which also have large convection cells. We plot the parallax uncertainties of our 47~Tuc sample stars as a function of $G$ magnitude in the middle panel of Fig.~\ref{fig:47Tuc-test}. The stars brightest in the $K$ band are highlighted as filled symbols in that panel. The four faintest stars in $G$ have relatively large parallax uncertainties, but three of them are also very bright in $K$ (i.e., they have a red $G-K$ colour). Unfortunately, as the sample stars span less than one magnitude in $G$, we cannot test very well \Gaia's performance as a function of $G$. As a consequence, our results do not allow one to identify the main intrinsic or extrinsic parameter responsible for increased parallax uncertainties in AGB stars compared to the bulk of the Gaia sources.

In addition, it is important to mention that there is a systematic bias to the Gaia parallaxes of stars located in dense regions of the sky, such as a globular cluster, due to spatial correlation. According to \citet{2021MNRAS.505.5978V}, a parallax uncertainty of at least 0.011 mas is more appropriate for stars in a globular cluster. However, uncertainties are -- in general -- larger for the 47~Tuc LPVs anyway. Also, \citet{2021MNRAS.506.2269E} found that the parallax uncertainty is increased for binaries with projected separations $\lesssim4\arcsec$. Therefore, we may expect that the parallax uncertainty is enhanced for stars close to the cluster centre, where the stellar density is highest. We plot the parallax uncertainty as a function of the angular distance from the cluster centre ($r$) in the lower panel of Fig.~\ref{fig:47Tuc-test}. Indeed, we see that the parallax uncertainty increases at $r\lesssim2\arcmin$. To demonstrate that the parallax uncertainty depends on both the $K$ band magnitude and the distance from the cluster centre, we plot the stars at $r\leq1\arcmin$ as filled symbols in the upper panel, and stars $K\leq6\fm5$ as filled symbols in the lower panel of Fig.~\ref{fig:47Tuc-test}.

Independent of the cause of the enhanced parallax uncertainties, we can perform a theoretical test if they are correctly estimated in \Gaia DR3. For this, we adopted the same methodology as \citet{2021MNRAS.506.2269E} to estimate the reliability of the parallax uncertainties from binary components and treat each pair of LPVs as components of a binary system. We can here neglect the spatial extension of 47~Tuc because the half-light radius is only $\sim2.9\arcmin$ \citep{1993ASPC...50..347T}, corresponding to about 3.7\,pc, whereas the distance to the cluster is $\sim4.45$\,kpc. Having $n=41$ stars thus gives us $n\cdot(n-1)/2=820$ stellar pairs $i,j$ ($i\ne j$) for which we calculated the quantity $\Delta\varpi/\sigma_{\varpi,{\rm tot}}=(\varpi_i - \varpi_j)/\sqrt{\sigma_{\varpi,i}^2+\sigma_{\varpi,j}^2}$. For perfectly estimated parallax uncertainties, $\Delta\varpi/\sigma_{\varpi,{\rm tot}}$ should follow a normal distribution with a width of 1.0. In fact, we find that $\Delta\varpi/\sigma_{\varpi,{\rm tot}}$ distributes with a width of 1.34. The ratio of the standard deviation of the parallaxes to the average parallax uncertainty gives the almost identical value of $0.0300/0.0222=1.35$. We may compare this with the parallax uncertainty inflation factors inferred from widely separated pairs by \citet[][their Eq.~16]{2021MNRAS.506.2269E}. Our 47~Tuc LPVs have \Gaia $G$ band magnitudes in the relatively narrow range between 10\fm71 and 11\fm59. \citet{2021MNRAS.506.2269E} estimate parallax uncertainty inflation factors between 1.11 and 1.16 for this magnitude range. Thus, the parallax uncertainties of the 47~Tuc LPVs are underestimated by a somewhat larger factor than would be expected from this empirical formula, even if all but one of the sample stars have {\tt ruwe}~$<1.4$.

\section{Stars with changing pulsation periods}

Several Miras and SRVs are known to have pulsation periods that change significantly on timescales of decades or centuries. \citet{2005AJ....130..776T} find that $\sim4$\% of their sample stars change periods at a level $>3\sigma$, and \citet{2023A&A...672A.165M} studied in more detail their amplitudes of period change. Continuous, sudden, and meandering period changes have been reported in the literature \citep{2002JAVSO..31....2Z,2011A&A...531A..88U}. Different mechanisms have been suggested to explain these period changes. The leading explanation is that continuous and sudden period changes are caused by a recent TP, and we see the stars in different phases of the TP \citep{1981ApJ...247..247W}. In this picture, a star experiencing a sudden decrease in period undergoes the onset of a TP, while a continuous change corresponds to a later phase of the TP cycle. A third dredge-up event in the aftermath of a TP may also cause a period change, if it happens to increase the atmospheric C/O ratio from $\lesssim1$ to $>1$ \citep{2016A&A...585A.145U}. \citet{2023A&A...672A.165M} found evidence that also the meandering period change could be connected to the effects of TPs.

\begin{figure}
\centering
\includegraphics[width=\linewidth]{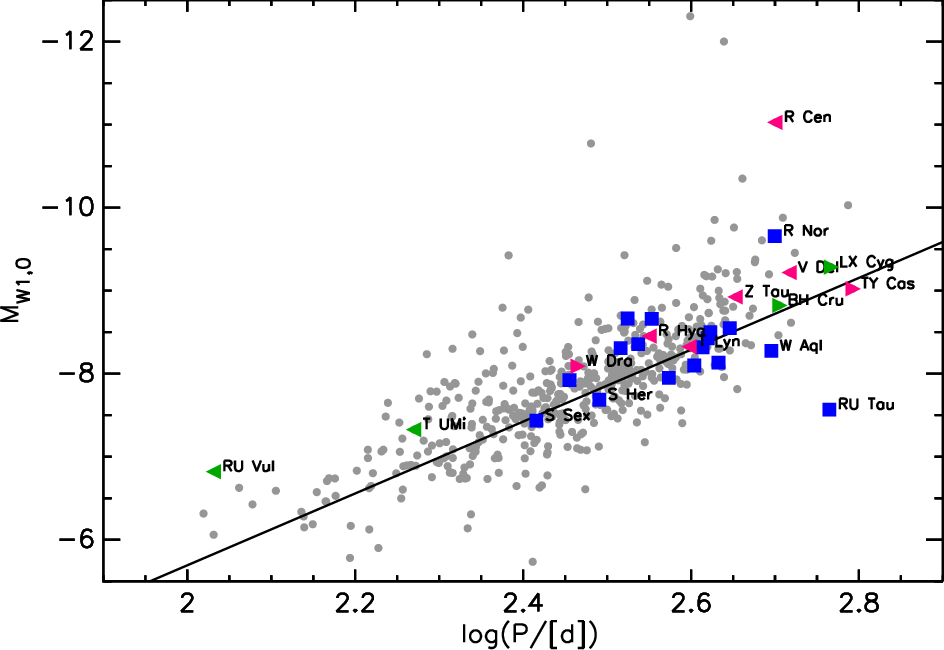}
\caption{Period--luminosity($W1$) diagram of all Karlsson Miras, with stars with changing pulsation periods highlighted. Magenta triangles represent Miras with continuously changing periods, with the orientation of the triangle indicating increasing (right-pointing) or decreasing (left-pointing) periods. Green triangles represent stars with sudden period changes, with the same meaning of the triangle orientation, and blue squares represent Miras with meandering periods. RU~Vul is the only nominal SRV, and it is not contained in the Karlsson sample. The grey dots represent all other Miras with relatively stable periods.}
\label{fig:changingP}
\end{figure}

So far, it is not known where Miras or SRVs undergoing such period changes are located in PL diagrams and whether they follow the PL relations of stars with more constant periods. The evolutionary models presented in Sect.~\ref{sec:PL-test} are not useful as a reference here because they do not well resolve the short phases around the TPs, and the model tracks switch to one of the overtone modes during the phases of fastest period change. Nevertheless, the data presented in this paper allowed us to address this question for the first time. Here, it is of particular importance to have contemporary (IR) photometric observations and period determinations. Figure~\ref{fig:changingP} shows the PL($W1$) diagram of all 510 Karlsson Miras with \Gaia parallax distances. Stars with changing pulsation periods identified in the literature (see the above-cited papers) are highlighted by coloured symbols. The three change types are identified by different symbols and colours. All symbols of stars with continuously and suddenly changing periods are labelled with their names, and a few with meandering periods are also labelled. Meandering-period stars not labelled in Fig.~\ref{fig:changingP} are RS~Aql, AF~Car, U~CMi, T~CMi, TY~Cyg, T~Hya, S~Ori, SS~Peg, RU~Sco, Z~Sco, T~Ser, and Z~Vel.

A few stars in Fig.~\ref{fig:changingP} do not follow the relation formed by the other stars well. One of them is R~Cen, a Mira whose period has continuously decreased from 570\,d in 1870 to $505-510$\,d in 2000 \citep{2001PASP..113..501H}. The list of T.\ Karlsson reports a current period of just over 500\,d. The star is thought to be an intermediate-mass star ($M\gtrsim4M_{\sun}$) undergoing hot bottom burning \citep[HBB;][]{2013A&A...555L...3G}. Such stars are expected to be shifted from the relation formed by lower-mass stars because of their higher mass and the additional luminosity generated by the HBB process \citep{2019A&A...631A..24L}. However, R~Cen is shifted from the relation by a large margin. R~Nor is a candidate for being an intermediate-mass AGB star undergoing HBB, too, but it is located above the relation by a smaller amount. The \Gaia parallax of R~Cen is also relatively uncertain, $\sigma_{\varpi}/\varpi\approx0.22$. We therefore suspect that its \Gaia distance is overestimated by a considerable amount.

RU~Tau has a meandering pulsation period and is also separated from the other Miras. However, it is located much below the relation formed by the other Miras, a locus that it does not share with any other star. Its \Gaia parallax is also relatively uncertain $\sigma_{\varpi}/\varpi\approx0.17$, and RU~Tau is at a large distance of $d\approx2.5$\,kpc from the Sun.

The only non-Mira included in Fig.~\ref{fig:changingP} is RU~Vul. It was originally classified as an SRV of subtype a, undergoing a sudden period decrease that commenced in 1954. We include it in Fig.~\ref{fig:changingP} with a period of 108\,d measured by \citet{2016AN....337..293U} from observations obtained between 2011 and 2013. RU~Vul is shifted from the FM PL($W1$) relation to shorter period and/or higher luminosity. \citet{2020MNRAS.491.1174M} showed that the star is swiftly becoming brighter in the mid-infrared as a result of rapid dust formation close to the star. However, this probably does not affect the $W1$ band, and the overall luminosity has probably decreased over the last decades \citep{2020MNRAS.491.1174M}. The shift to a shorter period could therefore be an indication that RU~Vul is evolving from the FM towards one of the overtone pulsation modes. Similarly, T~UMi has been shown by \citet{2019ApJ...879...62M} to have transitioned to a double-mode pulsation state with an FM period of just under 200\,d at the end of 2018 and a first overtone mode period of 110\,d, which would put it close to RU~Vul. We include it in Fig.~\ref{fig:changingP} with a period of $\sim190$\,d as listed by T.\ Karlsson, corresponding to the FM period. Thus, RU~Vul and T~UMi could be experiencing the same evolutionary phenomenon.

Apart from these few outliers, all other stars appear to follow the PL($W1$) relation relatively closely. Many of the stars with changing pulsation periods are actually contained in the gold sample defined above. Therefore, we conclude with some confidence that Mira stars with changing pulsation periods, in general, follow the FM PL($W1$) relation relatively closely. In turn, this means that the distances of Miras with changing pulsation periods can be inferred from their pulsation periods and apparent (IR) magnitudes with reasonable accuracy, provided that these quantities are observed at the same time and the star has not yet moved to overtone pulsation.

\section{Discussion and conclusions}\label{sec:discussion}

We collected high-quality well-curated data of solar neighbourhood Miras to construct their PL relations in nine near-IR bands, most of them for the first time. Our database was built from two multi-epoch IR catalogues, namely the COBE/DIRBE catalogue of variable stars of \citet{2010ApJS..190..203P} and the WISE/unTimely catalogue of \citet{2023AJ....165...36M}. The pulsation periods of the stars were adopted from \citet{2013JAVSO..41..348K}, who determined them from contemporaneous optical observations. These data were combined with \Gaia DR3 parallax distances derived and published by \citet{2021AJ....161..147B}. The stars in the gold sample, defined to have relative parallax uncertainties of $\le10$\%, form well-defined PL relations without sigma-clipping or further removal of outliers, except for one temporarily obscured C star (Sect.~\ref{sec:PLDs}). Comparison with data from the literature of the PL relations of (O-rich) Miras in the more metal-poor LMC confirms that the Galactic Miras are fainter than the LMC Miras by $0\fm16-0\fm36$ in those bands and that the PL relations of Galactic Miras are steeper than those of LMC Miras (Sect.~\ref{sec:PLR}).

In Sect.~\ref{sec:synthSEDs}, we constructed synthetic SEDs based on the nine PL relations established before. We fitted the SEDs at six $\log(P)$ values representative of our sample stars with a blackbody curve. The WISE bands were excluded from the fit because they significantly deviated from the Planck curve, indicating remaining issues with reconstructing the flux of these bright, partly saturated sources. From the blackbody fit, we derived the period--temperature, period--bolometric-luminosity, and period--radius relations (see Eqs.~\ref{Eq:Tbb}-\ref{Eq:Rstar}). They may be useful as empirical relationships for other studies of AGB variables.

We also used our data to investigate whether the parallax uncertainties of Mira stars given in \Gaia DR3 are underestimated. Mira variables are very extended stars with low surface gravity and are thought to have large convective cells on their photospheres, resulting in significant surface brightness variations. These variations may impact the position measurements used for the parallax determination. The literature suggests that the parallax uncertainties of AGB stars in general and Mira variables in particular are significantly underestimated \citep{2018evn..confE..43V,2022A&A...667A..74A}. In this work, we used three different approaches to test the \Gaia parallax uncertainties. The first one (Sect.~\ref{sec:PL-test}) uses the observed scatter of stars around the PL($K$) relation. If the true parallax uncertainties were considerably larger than the one quoted in the \Gaia catalogue, we may expect to observe an ill-defined PL relation of Miras. We applied a small grid of \texttt{COLIBRI} evolutionary tracks to model the intrinsic width of the FM PL relation for a solar neighbourhood sample due to stellar evolution on the AGB and the range of initial masses. The result of this test is that the EIF to be applied to the parallax uncertainties is consistent with 1.0, which means that the original \Gaia uncertainties are realistic. An even better agreement between observed and modelled scatter is reached for an EIF of 1.3, and an EIF of 1.7 marginally agrees with the observations. This test is not highly stringent, and we discussed the shortcomings of the evolutionary grid used to model the scatter. Nevertheless, it indicates that large EIFs ($>2$), as reported in the literature, are unlikely.

The second test (Sect.~\ref{sec:VLBI-test}) compared the \Gaia parallaxes and their uncertainties to the respective measurements derived from VLBI observations. Before applying this test, we improved on the pulsation periods of the stars in the common sample. With the improved periods, the Miras form a very well defined PL relation in the WISE $W1$ band that extends to the longest periods ($P\gtrsim1000$\,d). This indicates that long-period OH/IR stars follow the same near-IR PL relation as shorter-period Miras, as long as circumstellar dust does not significantly diminish the light in the chosen photometric band. Restricting the common sample to stars that have relative parallax uncertainties $\le10$\% in both catalogues and removing two further significant outliers reduces the EIF of the \Gaia parallax uncertainties to $\sim2.3$ (Table~\ref{tab:EIFs}). However, for a faint sample, the EIF is reduced to $\sim1.8$. These factors apply only if one assumes that the VLBI parallax uncertainties are reliable and do not require an inflation factor. \citet{2020PASJ...72...50V} conceded that the VLBI parallax error analysis differs from paper to paper and that the formal uncertainty could underestimate the error in the averaged parallax value if systematic errors from atmospheric phase fluctuation are not fully taken into account. Thus, rather than the one-sided EIF, the two-sided EIF estimated in this work might be more applicable. We also caution that this test relies on a relatively small sample of stars with VLBI parallaxes because they are laborious to measure.

The third and final test was applied to a sample of 41 LPVs on the AGB of the globular cluster 47~Tuc. This sample of stars can be considered to be located at an identical distance from the Sun, and if their parallax uncertainties are estimated correctly, their spread in measured distances and combined parallax uncertainties should agree in a statistical sense. This test is the most stringent of the three and results in an EIF of 1.34. The disadvantage of this test, however, is that the sample stars are relatively low-mass and metal-poor; thus, they are not as extended as the more massive and metal-rich solar neighbourhood Miras.

All three tests suggest EIFs of the \Gaia parallax uncertainties of AGB stars in the range of 1.0 to 1.8. Even though it would be possible to combine the results in a single number, we refrained from doing so because the methods and the samples to estimate the EIF are so different. Our results may be compared to other EIF estimates from the literature. \citet{2021MNRAS.506.2269E} analysed resolved binaries in the \Gaia catalogue and found a peak EIF of 1.3 at $G=13$. This peak is at a fainter $G$ band magnitude than most of our Miras are on average. \citet{2023MNRAS.523.2369S}, on the other hand, also analysed Galactic Miras and their spread about the PL relation with a probabilistic model. Their result suggests that for five-parameter solutions, the EIF is around 1.3 for brighter ($G\sim9$) and fainter ($G\sim16$) sources, but for more intermediate ($G\sim12$), the factor increases to around 1.6. Very recently, \citet{2026Galax..14...26N} presented a detailed analysis of \Gaia and VLBI parallaxes of 43 Galactic LPVs and found a two-sided EIF ($\xi$ in our notation) of 1.43. We thus conclude that our findings agree with recent results from the literature. Significantly larger EIFs can probably be ruled out with high confidence, but more tests on the reliability of the \Gaia parallax uncertainties of AGB stars are highly encouraged.

As one can appreciate from Figs.~\ref{fig:Gaia-VLBI_PL-W1} and \ref{fig:changingP}, stars with relatively uncertain parallaxes ($\sigma_{\varpi}/\varpi>0.1$) tend to be located above the PL relation formed by the gold sample stars with precise parallaxes. We confirmed this by inspecting the location of stars with $\sigma_{\varpi}/\varpi>0.1$ in the PL([2.2]) and PL($W1$) diagrams. On average, the stars are above the best-fit line by $\sim0\fm6$ and $\sim0\fm3$, respectively. {The systematic offset from the PL relation suggests} that the distances of AGB stars with larger uncertainties tend to be overestimated in the catalogue of \citet{2021AJ....161..147B}, even though their probabilistic approach should take into account asymmetric probability distributions. However, as the range of deviation from the PL relations is large, it is difficult to derive a general law to improve uncertain parallax distance estimates of Miras. Finally, we inspected the location of Miras with changing pulsation periods in the PL($W1$) diagram. We showed that they follow the FM PL($W1$) relation formed by other stars relatively closely, except for a few outliers. Thus, their distances can be inferred from their pulsation periods and apparent IR magnitudes with reasonable accuracy if periods and mean magnitudes are measured contemporaneously.

\begin{acknowledgements}
We thank Aaron Meisner for help with the WISE/unTimely photometry, {and the anonymous referee for comments that helped to improve the paper}. This research was funded in part by the Austrian Science Fund (FWF) 10.55776/F8100. For open access purposes, the author has applied a CC BY public copyright license to any author accepted manuscript version arising from this submission. We acknowledge with thanks the variable star observations from the AAVSO International Database contributed by observers worldwide and used in this research. This work has made use of data from the European Space Agency (ESA) mission \Gaia (\url{https://www.cosmos.esa.int/gaia}), processed by the \Gaia Data Processing and Analysis Consortium (DPAC, \url{https://www.cosmos.esa.int/web/gaia/dpac/consortium}). Funding for the DPAC has been provided by national institutions, in particular the institutions participating in the \Gaia Multilateral Agreement. This publication makes use of data products from the Two Micron All Sky Survey, which is a joint project of the University of Massachusetts and the Infrared Processing and Analysis Center/California Institute of Technology, funded by the National Aeronautics and Space Administration and the National Science Foundation. This publication makes use of data products from the Wide-field Infrared Survey Explorer, which is a joint project of the University of California, Los Angeles, and the Jet Propulsion Laboratory/California Institute of Technology, funded by the National Aeronautics and Space Administration.
\end{acknowledgements}

\bibliographystyle{aa}
\bibliography{References}

\end{document}